# A new view on quantum electrodynamics

V. A. Golovko

Moscow Polytechnic University

Bolshaya Semenovskaya 38, Moscow 107023, Russia

E-mail: fizika.mgvmi@mail.ru

Abstract

We analyze the equations of quantum electrodynamics and establish that the electron must be described by two bispinors that satisfy two mutually connected Dirac equations. The equations of the electronic and electromagnetic fields are reformulated in terms of *c*-numbers, which enables one to elucidate the structure of the electron. Although the equations obtained allow only for numerical solution, some characteristics of the electron, in particular its size, can be found at this stage. It is shown also that the Dirac equation should, instead of the mass of the electron, contain a combination involving the electron Compton wavelength. In this case the equations obtained can be used not only for the description of the electron but also for the description of other leptons, which will allow one to find the mass spectrum of leptons.

## 1. Introduction

It was shown in Ref. [1] that all effects and phenomena that seemingly confirm the corpuscular nature of light (the photoelectric effect, the Compton effect etc.) can be explained on the basis of the undulatory theory of light. The unique mechanism that explains all these effects and phenomena is resonance interaction of the incident electromagnetic wave with an atom or another resonator wherein an essential role is played by reverse impact of the resonator on the incident wave. In case the resonance occurs, the resonator distorts the incident wave efficaciously and rapidly sucking in its energy. As a result, the amplitude of the wave in the vicinity of the resonator substantially grows producing the relevant action on the resonator. All of this looks like a particle (a photon) hit the resonator although there are no photons in nature.

At the same time the photon is one of the basic notions in quantum electrodynamics (QED). It should be emphasized at once that the photon in QED is by no means a tiny particle: the term "photon" implies an elementary excitation of an electromagnetic field in the form of a plane wave with energy $\hbar\omega$, the excitation that concerns the whole of the volume occupied by the system in question. It is shown in Ref. [1] that the quantization of the electromagnetic field in QED signifies a preliminary breakdown of the initial electromagnetic field into elementary plane waves, each of which having an energy of $\hbar\omega$ that can be efficaciously absorbed in the resonance interaction. One endows the elementary plane waves with properties of excitations of a harmonic oscillator. As a result, the quantization of the electromagnetic field in QED is merely a convenient mathematical tool that enables one to consider the result of a process without going into detail of the process itself. At the same time, the electronic field is also quantized in QED. In the first part of the present paper it will be shown that the quantization of the electronic field is a convenient mathematical tool as well although it has another character than the quantization of the electromagnetic field. An important result of the first part is the fact that the electron must be described by two bispinors rather than by one.

The Dirac equation employed in QED is regarded as an equation of motion for the electron in an external field. On a basis of that equation, however, one deduces properties of the electron such as the spin and the magnetic moment, the later with an anomalous part calculated very exactly. This can occur only if the Dirac equation, being the equation of motion for the electron, simultaneously describes the structure of the electron. One meets with an analogical situation in general relativity whose equations are simultaneously equations of motion and equations of the structure of the gravitational field. To find the structure of the electron in QED is, however, impossible inasmuch as the wave function that should describe the structure is an operator (a *q*- number) rather than a usual (classical) number (a *c*-number). Therefore the challenge arises to reformulate the equations of QED in terms of *c*-numbers.

In Sec. 2 of the present paper, the equations of QED are revisited and reduced to a form in which *q*-numbers can be replaced by *c*-numbers. It emerges also that the electron must be described by two bispinors $\psi_1$ and $\psi_2$. The meaning of the second quantization in QED is discussed as well. In Sec. 3 the equations for the bispinors and electromagnetic field are deduced from the principle of least action. The knowledge of the Lagrangian enables one to find the energy-momentum tensor and the three-dimensional angular momentum vector as well. Regular and localized in space (particle-like) solutions to the equations obtained that should correspond to the electron are considered in Sec. 4. The initial equations whose number is 16 are reduced to a system of 6 equations for 6 basic functions. Such a system can be solved only numerically. We discuss those properties of the electron, in particular its size, that can be found at this stage. In Sec. 5 are analyzed the fundamental constants that enter into the Dirac equation, and it is shown that the equation must, instead of the mass of the electron, contain a combination involving the electron Compton wavelength. In this case the equations obtained can be applied not only for the description of the electron but also for the description of other leptons. The results obtained are discussed in the concluding section 6.

## 2. Analysis of equations of quantum electrodynamics

Let us write down the principal equations of QED (see, e. g., Ref. [2], § 102). The operators of the electronic and electromagnetic fields $\widehat{\psi}(x)$ and $\widehat{A}_\mu(x)$ satisfy the Dirac and Maxwell equations (we mark operators with an arc at the top):

$$i\gamma^\mu \frac{\partial\widehat{\psi}}{\partial x^\mu} - e\widehat{A}_\mu\gamma^\mu\widehat{\psi} - m\widehat{\psi} = 0\,, \tag{2.1}$$

$$\frac{\partial \widehat{F}^{\mu\nu}}{\partial x^\nu} = -4\pi e\widehat{j}^{\,\mu}\,, \tag{2.2}$$

where $\widehat{F}^{\mu\nu} = \partial\widehat{A}^\nu / \partial x_\mu - \partial\widehat{A}^\mu / \partial x_\nu$ is the electromagnetic field tensor, and the particle current density four-vector $\widehat{j}^{\,\mu}$ is

$$\widehat{j}^{\,\mu} = \widehat{\overline{\psi}}\gamma^\mu\widehat{\psi}\,. \tag{2.3}$$

Here and henceforth, if the opposite is not pointed out, we utilize the definitions and notation adopted in [2].

As is usual, we expand the operator $\widehat{\psi}$ in terms of the eigenfunctions $\psi_s$ of a complete set of possible states of a free particle:

$$\hat{\psi}(x)=\sum_s \hat{a}_s\psi_s^{(+)}(x)+\sum_s \hat{b}_s^{+}\psi_s^{(-)}(x)\,, \tag{2.4}$$

wherein $\hat{a}_s$ and $\hat{b}_s^{+}$ are the annihilation and creation operators of particles respectively. It will be noted for what follows that

$$\psi_s^{(+)}\propto e^{-i\omega_s t},\quad \psi_s^{(-)}\propto e^{i\omega_s t}\,, \tag{2.5}$$

where $\omega_s = |\omega_s|$ while $\omega_s \geq m$ ( see also Ref. [3]). In the following we shall speak either about energies $\varepsilon$ or about frequencies $\omega$ seeing that $\varepsilon = \omega$ once $\hbar = 1$. For the sake of convenience we rewrite the expansion of (2.4) in the form

$$\hat{\psi}=\hat{\psi}_a+\hat{\psi}_b^c;\quad \hat{\psi}_a=\sum_s \hat{a}_s\psi_s^{(+)}(x),\ \ \hat{\psi}_b^c=\sum_s \hat{b}_s^{+}\psi_s^{(-)}(x)\,. \tag{2.6}$$

Here and henceforth, the superscript $c$ signifies charge conjugation:

$$\hat{\psi}^c=C\hat{\bar{\psi}}=-\hat{\bar{\psi}}C,\ \ \hat{\bar{\psi}}^c=C^{-1}\hat{\psi}=-\hat{\psi}C^{-1}\,, \tag{2.7}$$

where $C$ is the charge-conjugation matrix. When deriving the second expressions in (2.7) account is taken of the fact that $\gamma^\mu C=-C\tilde{\gamma}^\mu$ ($C \equiv U_C$ from [2]). We write down the Dirac conjugate operator as well:

$$\hat{\bar{\psi}}=\hat{\bar{\psi}}_a+\hat{\bar{\psi}}_b^c;\quad \hat{\bar{\psi}}_a=\sum_s \hat{a}_s^{+}\bar{\psi}_s^{(+)}(x),\ \ \hat{\bar{\psi}}_b^c=\sum_s \hat{b}_s\bar{\psi}_s^{(-)}(x)\,. \tag{2.8}$$

Here instead of (2.5) we have

$$\bar{\psi}_s^{(+)}\propto e^{i\omega_s t},\quad \bar{\psi}_s^{(-)}\propto e^{-i\omega_s t}\,. \tag{2.9}$$

The commutation relations between the operators can be obtained from (2.6) and (2.8) with account taken of the fact that the operators $\hat{a}_s$ and $\hat{b}_s$ satisfy the anticommutation rules

$$\left\{\hat{a}_s,\hat{a}_{s'}^{+}\right\}_+=\delta_{ss'},\quad \left\{\hat{b}_s,\hat{b}_{s'}^{+}\right\}_+=\delta_{ss'}\,, \tag{2.10}$$

and all other pairs of the operators $\hat{a}_s,\hat{a}_s^{+},\hat{b}_s,\hat{b}_s^{+}$ anticommute. We are in need of the following commutation relations

$$\left\{\hat{\bar{\psi}}_{b,i}^c(x),\hat{\psi}_{b,k}^c(x')\right\}_+=S_{ik}(x,x'),\quad S_{ik}(x,x')=S_{ki}(x',x)=\sum_s \bar{\psi}_{s,i}^{(-)}(x)\psi_{s,k}^{(-)}(x')\,, \tag{2.11}$$

$$\left\{\hat{\bar{\psi}}_{b,i}^c(x),\,\hat{\psi}_{a,k}(x')\right\}_+=0\,, \tag{2.12}$$

where we write down explicitly, after the comma, the bispinor indices $i$, $k$ = 1, 2, 3, 4. With use made of (2.7) it follows therefrom that

$$\hat{\bar{\psi}}_b^c(x)\gamma^\mu\hat{\psi}_b^c(x)=-\hat{\bar{\psi}}_b\gamma^\mu\hat{\psi}_b+\mathrm{tr}\left(\gamma^\mu S\right), \tag{2.13}$$

$$\hat{\bar{\psi}}_b^c(x)\gamma^\mu\hat{\psi}_a(x)=\hat{\psi}_a C^{-1}\gamma^\mu\hat{\psi}_b\,. \tag{2.14}$$

It may be remarked that Eq. (2.13) without the last term is adduced in Ref. [2], § 28.

We substitute the expansions of (2.6) and (2.8) into the current density of (2.3) with account taken of (2.13) and (2.14). According to the common practice in QED we neglect the last term in (2.13) that does not depend upon the state of the field, which amounts to saying that the term is relevant to vacuum (usually, such terms are infinite). The result can be recast in the form

$$\hat{j}^{\mu} = \hat{j}^{\mu}_{(0)} + \hat{j}^{\mu}_{(1)} + \hat{j}^{\mu}_{(2)}, \tag{2.15}$$

wherein

$$\hat{j}^{\mu}_{(0)} = \bar{\hat{\psi}}_a \gamma^{\mu} \hat{\psi}_a - \bar{\hat{\psi}}_b \gamma^{\mu} \hat{\psi}_b, \quad \hat{j}^{\mu}_{(1)} = \hat{\psi}_a C^{-1} \gamma^{\mu} \hat{\psi}_b, \quad \hat{j}^{\mu}_{(2)} = \bar{\hat{\psi}}_a \gamma^{\mu} C \bar{\hat{\psi}}_b. \tag{2.16}$$

Accordingly, in view of (2.2) we shall have for the components of the four-potential that

$$\hat{A}_{\mu} = \hat{A}^{(0)}_{\mu} + \hat{A}^{(1)}_{\mu} + \hat{A}^{(2)}_{\mu}. \tag{2.17}$$

With use made of (2.7) and of properties of the matrices $C$ and $\gamma^{\mu}$ we find that

$$\left(\hat{j}^{\mu}_{(1)}\right)^{*} = \left(\hat{\psi}_a C^{-1} \gamma^{\mu} \hat{\psi}_b\right)^{*} = -\hat{j}^{\mu}_{(2)}. \tag{2.18}$$

It follows therefrom that the sum $\hat{j}^{\mu}_{(1)} + \hat{j}^{\mu}_{(2)} = \hat{j}^{\mu}_{(1)} - \left(\hat{j}^{\mu}_{(1)}\right)^{*}$ is a purely imaginary quantity. As long as the current cannot be imaginary, we are forced to put $\hat{j}^{\mu}_{(1)} + \hat{j}^{\mu}_{(2)} = 0$, which entails

$$\hat{\psi}_a C^{-1} \gamma^{\mu} \hat{\psi}_b + \bar{\hat{\psi}}_a \gamma^{\mu} C \bar{\hat{\psi}}_b = 0. \tag{2.19}$$

We can arrive at this relation in another way. According to (2.5) and (2.9), $\hat{j}^{\mu}_{(1)}$ contains summands with the time factors $e^{-2i\omega_s t}$ while $\hat{j}^{\mu}_{(2)}$ with the time factors $e^{2i\omega_s t}$, which corresponds to transitions between states of positive energy and those of negative energy that are separated by an energy interval of $2m$ (recall that $\omega_s \geq m$). According to the prescriptions of QED such transitions are to be forbidden, which leads to (2.19).

Along with $\hat{j}^{\mu}_{(1)} + \hat{j}^{\mu}_{(2)}$, the sum $\hat{A}^{(1)}_{\mu} + \hat{A}^{(2)}_{\mu}$ must be imaginary, that is to say,

$$\left(\hat{A}^{(1)}_{\mu} + \hat{A}^{(2)}_{\mu}\right)^{*} = -\hat{A}^{(1)}_{\mu} - \hat{A}^{(2)}_{\mu}. \tag{2.20}$$

It should be emphasized that we cannot neglect $\hat{A}^{(1)}_{\mu}$ and $\hat{A}^{(2)}_{\mu}$ separately although the imaginary sum $\hat{A}^{(1)}_{\mu} + \hat{A}^{(2)}_{\mu}$ will be thrown away in what follows.

We turn now to Eq. (2.1). Only solutions of positive energy have a physical meaning. As usual we presume that the positive energies (frequencies) correspond to the factor $e^{-i\omega_s t}$ as in the first term in (2.4) and (2.6). We place (2.6) and (2.17) in (2.1) and delete terms with negative frequencies for a separate equation that such terms produce is of no interest to us. In so doing we take into consideration (2.5) and the fact that $\hat{A}^{(0)}_{\mu}$ as $\hat{j}^{\mu}_{(0)}$ of (2.16) contains low frequencies, which amounts to saying that products, for example, of the type $\hat{A}^{(0)}_{\mu} e^{-i\omega_s t}$ are to be assigned to terms with positive frequencies. From what was said about $\hat{j}^{\mu}_{(1)}$ and $\hat{j}^{\mu}_{(2)}$ it follows that the quantity $\hat{A}^{(1)}_{\mu}$ contains terms with the time factors $e^{-2i\omega_s t}$ while $\hat{A}^{(2)}_{\mu}$ with $e^{2i\omega_s t}$. As a consequence we obtain the equation

$$i\gamma^{\mu}\frac{\partial\hat{\psi}_a}{\partial x^{\mu}} - e\hat{A}^{(0)}_{\mu}\gamma^{\mu}\hat{\psi}_a - m\hat{\psi}_a - e\hat{A}^{(1)}_{\mu}\gamma^{\mu}\hat{\psi}_a - e\hat{A}^{(1)}_{\mu}\gamma^{\mu}\hat{\psi}^{c}_{b} = 0\,. \tag{2.21}$$

The positive frequencies figure also in the second term in (2.8). For this reason, we write down the complex conjugate equation with respect to (2.1) having regard to the fact that our four-potential $\hat{A}_{\mu}$ is complex:

$$i\frac{\partial\hat{\bar{\psi}}}{\partial x^{\mu}}\gamma^{\mu} + e\hat{A}^{*}_{\mu}\hat{\bar{\psi}}\gamma^{\mu} + m\hat{\bar{\psi}} = 0\,. \tag{2.22}$$

We substitute (2.8) and (2.17) here and take (2.20) into account. Retaining terms with positive frequencies alone yields

$$i\frac{\partial\hat{\bar{\psi}}^{c}_{b}}{\partial x^{\mu}}\gamma^{\mu} + e\hat{A}^{(0)}_{\mu}\hat{\bar{\psi}}^{c}_{b}\gamma^{\mu} + m\hat{\bar{\psi}}^{c}_{b} - e\hat{A}^{(1)}_{\mu}\hat{\bar{\psi}}^{c}_{b}\gamma^{\mu} - e\hat{A}^{(1)}_{\mu}\hat{\bar{\psi}}_{a}\gamma^{\mu} = 0\,. \tag{2.23}$$

In Eqs. (2.21) and (2.23), the fourth terms contain summands with the time factors $e^{-3i\omega_s t}$. Characteristic of QED is the appearance of ultraviolet divergences in various integrals at high energies. To eliminate the divergences one resorts, in one way or another, to a high-energy cutoff. In our case the cutoff corresponds to discarding the fourth terms in Eqs. (2.21) and (2.23).

It is convenient to introduce $\hat{\psi}_1$ and $\hat{\psi}_2$ in place of the bispinors $\hat{\psi}_a$ and $\hat{\psi}_b$ according to

$$\hat{\psi}_1 = \hat{\psi}_a,\ \ \hat{\psi}_2 = iC\hat{\bar{\psi}}_b;\quad \psi_b = iC\overline{\psi}_2,\ \ \hat{\bar{\psi}}^{c}_{b} = i\hat{\bar{\psi}}_2\,. \tag{2.24}$$

Besides, instead of the purely imaginary operator $\hat{A}^{(1)}_{\mu}$ we introduce a real operator $\hat{v}_{\mu}$ according to $\hat{A}^{(1)}_{\mu} = i\hat{v}_{\mu}$ and replace $\hat{A}^{(0)}_{\mu}$ by $\hat{A}_{\mu}$ in view of (2.17) seeing that $\hat{A}^{(1)}_{\mu} + \hat{A}^{(2)}_{\mu} = 0$. Then Eq. (2.21) assumes the form

$$i\gamma^{\mu}\frac{\partial\hat{\psi}_1}{\partial x^{\mu}} - e\hat{A}_{\mu}\gamma^{\mu}\hat{\psi}_1 - m\hat{\psi}_1 - e\hat{v}_{\mu}\gamma^{\mu}\hat{\psi}_2 = 0\,. \tag{2.25}$$

If one takes the complex conjugate of (2.23), the result can be represented as

$$i\gamma^{\mu}\frac{\partial\hat{\psi}_2}{\partial x^{\mu}} - e\hat{A}_{\mu}\gamma^{\mu}\hat{\psi}_2 - m\hat{\psi}_2 + e\hat{v}_{\mu}\gamma^{\mu}\hat{\psi}_1 = 0\,. \tag{2.26}$$

With account taken of (2.16) and of the fact that $\hat{j}^{\mu}_{(1)} + \hat{j}^{\mu}_{(2)} = 0$, Eq. (2.2) becomes

$$\frac{\partial\hat{F}^{\mu\nu}}{\partial x^{\nu}} = -4\pi e(\hat{\bar{\psi}}_1\gamma^{\mu}\hat{\psi}_1 - \hat{\bar{\psi}}_2\gamma^{\mu}\hat{\psi}_2)\,. \tag{2.27}$$

As to the condition of (2.19) it now has the form

$$\bar{\hat{\psi}}_1\gamma^\mu\hat{\psi}_2 + \bar{\hat{\psi}}_2\gamma^\mu\hat{\psi}_1 = 0\,. \qquad (2.28)$$

The four equations (2.25)– (2.28) constitute a closed system of equations for four quantities $\hat{\psi}_1$, $\hat{\psi}_2$, $\hat{A}_\mu$ and $\hat{v}_\mu$, and, what is important, all of them correspond to positive energies while negative energies that have no physical meaning do not enter into them. Besides, all prescriptions of QED are satisfied. Now, however, there is no need to consider these quantities to be operators ($q$-numbers), they can be regarded as $c$-numbers $\psi_1$, $\psi_2$, $A_\mu$ and $v_\mu$. In support of this conjecture we shall, independently, derive exactly the same equations for $\psi_1$, $\psi_2$, $A_\mu$ and $v_\mu$ in the next section.

It was noted in Introduction that the quantization of the electromagnetic field in QED is merely a convenient mathematical tool that enables one to break the electromagnetic wave up in advance into portions that can be efficaciously absorbed or emitted by the quantum system as a consequence of the resonance interaction. We see now that the quantization of the electronic field is in actual fact a mathematical instrument as well. This instrument permits one to obtain, through introduction of anticommuting $q$-numbers, correct signs in some formulae, for instance in Eqs. (2.13) and (2.14). But the utilization of $q$-numbers has another helpful aspect as well. By analogy with the fact that use of imaginary numbers absent in nature essentially extends the capabilities of mathematics the utilization of $q$-numbers enhances mathematical means as well. Let us give a case in point. In Ref. [4], § 42, it is shown that the solution to the d'Alembert and Dirac equations can be expressed in terms of $c$-numbers as well as in terms of $q$-numbers. The advantages of the second type of the solution manifest themselves when solving the equations simultaneously, the solutions being automatically expanded in powers of interaction whereas the whole apparatus of QED consists just in obtaining simultaneous solutions to the above equations with the help of expansions in powers of interaction (in powers of the fine-structure constant $\alpha = e^2/\hbar c$). At the same time, the utilization of $q$-numbers has a negative facet, too. It allows one to obtain physically rational results only if it becomes possible to express them finally in terms of $c$-numbers. In particular, the use of $q$-numbers gives no way of elucidating the structure of the electron inasmuch as the density distribution in the electron is expressible only in terms of $\hat{\psi}_1$ and $\hat{\psi}_2$.

We turn now to the question as to why characteristic of QED is the appearance of various divergences. The sole method for solving equations in the standard QED is expansion in powers of $\alpha$. However, in a zeroth approximation at $\alpha = 0$, i.e. at $e = 0$, the Dirac equation has no solutions bounded in space whereas the electron naturally is bounded in space. Thereupon one looks for corrections to the zeroth approximation that are proportional to powers of $\alpha$. Such corrections cannot, of course, change the divergent character of the zeroth approximation.

Although one succeeds in extracting information from divergent expressions with use made of various techniques, the divergence of the zeroth approximation manifests itself. Besides, not all of the characteristics of the electron can be calculated in this way. In particular, it is impossible to elucidate the structure of the electron as was mentioned above. One could circumvent the appearance of the divergences, provided that one does not resort to the expansions in powers of $\alpha$; but the apparatus of $q$-numbers in QED was formulated only within the framework of a perturbation theory.

## 3. New equations of quantum electrodynamics

Henceforth, all quantities will be considered to be $c$-numbers. The most important finding of the above investigation is that the electron must be described by two bispinors $\psi_1$ and $\psi_2$. The fact that not all of the electron properties can be described by one bispinor $\psi$ is clear from the following. From the Dirac equation it follows the equation of continuity $\partial j^\mu/\partial x^\mu = 0$ for the current density $j^\mu$ (2.3). Integrating the equation over all space yields

$$\int\limits_{(\infty)} \psi^* \psi dV = I \,, \tag{3.1}$$

where $I$ = constant. This equation is used for normalization $I = 1$. It should be emphasized that Eq. (3.1) is valid if the electron is moving in any field. If, however, the electron collides with a positron, there occurs annihilation and the electronic field has to disappear: $\psi = 0$. But this would be impossible in view of (3.1).

Once the electron is described by two bispinors $\psi_1$ and $\psi_2$, there must exist a relation between them because QED gets by with only one bispinor $\hat{\psi}(x)$. We can try and establish the relation from general considerations. As long as the bispinors $\psi_1$ and $\psi_2$ each have four components, the relation must consist of four equalities. The relation should be symmetric in $\psi_1$ and $\psi_2$ and should, in the simplest case, be bilinear in $\psi_1$ and $\psi_2$ (the existence of a complicated relation would be rather strange). In addition, the relation must be relativistically and gauge invariant. All of this leads uniquely to Eq. (2.28) which in the present notation will be

$$\overline{\psi}_1 \gamma^\mu \psi_2 + \overline{\psi}_2 \gamma^\mu \psi_1 = 0 \,, \tag{3.2}$$

where $\mu$ = 0, 1, 2, 3. Equation (3.2) may be treated as a condition of the absence of the “crossing” current since the current density in the theory of Dirac’s equation is defined by an expression of the type $j^\mu = \overline{\psi} \gamma^\mu \psi$.

In order to obtain equations of motion for all functions we need write down the Lagrangian $L(x)$. The terms in the Lagrangian describing the electromagnetic and electronic fields and their interaction are standard. Care should be taken only to choose properly the signs in front of the terms in accord with the equations of the preceding section. As long as there are the constraint relations of (3.2) between the functions, we are to resort to the method of Lagrange multipliers $\lambda_\mu$ which will be conveniently written as $\lambda_\mu = -ev_\mu$. Besides, we shall not put $\hbar = c = 1$ in what follows because explicitly writing down these constants we do not essentially clutter up formulae whereas this underlines the physical meaning of occurring terms. As a result we have

$$L(x) = -\frac{1}{16\pi} F_{\mu\nu} F^{\mu\nu} + \frac{ic\hbar}{2}\left( \overline{\psi}_1 \gamma^\mu \frac{\partial \psi_1}{\partial x^\mu} - \frac{\partial \overline{\psi}_1}{\partial x^\mu} \gamma^\mu \psi_1 - \overline{\psi}_2 \gamma^\mu \frac{\partial \psi_2}{\partial x^\mu} + \frac{\partial \overline{\psi}_2}{\partial x^\mu} \gamma^\mu \psi_2 \right)$$

$$- eA_\mu (\overline{\psi}_1 \gamma^\mu \psi_1 - \overline{\psi}_2 \gamma^\mu \psi_2) - mc^2 (\overline{\psi}_1 \psi_1 - \overline{\psi}_2 \psi_2) - ev_\mu (\overline{\psi}_1 \gamma^\mu \psi_2 + \overline{\psi}_2 \gamma^\mu \psi_1)\,. \tag{3.3}$$

First, we discuss the gauge invariance of the Lagrangian. As long as the function added to the four-potential $A_\mu$ under the gauge transformation is real while $v_\mu$ has originated from the imaginary part of $A_\mu$, the quantity $v_\mu$ should not change under the gauge transformation. Now, provided the bispinors $\psi_1$ and$\psi_2$ are transformed identically under the gauge, the Lagrangian of (3.3) remains gauge invariant. The Lagrangian is relativistically invariant as well on condition that $v_\mu$ is a four-vector.

Varying $L(x)$ with respect to $\overline{\psi}_1$, $\overline{\psi}_2$, $\psi_1$ and $\psi_2$ we obtain the equations of motion

$$ic\hbar\gamma^\mu \frac{\partial \psi_1}{\partial x^\mu} - eA_\mu \gamma^\mu \psi_1 - mc^2 \psi_1 - ev_\mu \gamma^\mu \psi_2 = 0\,, \tag{3.4}$$

$$ic\hbar\gamma^\mu \frac{\partial \psi_2}{\partial x^\mu} - eA_\mu \gamma^\mu \psi_2 - mc^2 \psi_2 + ev_\mu \gamma^\mu \psi_1 = 0\,, \tag{3.5}$$

$$ic\hbar \frac{\partial \overline{\psi}_1}{\partial x^\mu} \gamma^\mu + eA_\mu \overline{\psi}_1 \gamma^\mu + mc^2 \overline{\psi}_1 + ev_\mu \overline{\psi}_2 \gamma^\mu = 0\,, \tag{3.6}$$

$$ic\hbar \frac{\partial \overline{\psi}_2}{\partial x^\mu} \gamma^\mu + eA_\mu \overline{\psi}_2 \gamma^\mu + mc^2 \overline{\psi}_2 - ev_\mu \overline{\psi}_1 \gamma^\mu = 0\,. \tag{3.7}$$

The first two equations conform with (2.25) and (2.26) while the second two are the complex conjugates of the first two. Varying $A_\mu$ in $L(x)$ we have

$$\frac{\partial F^{\mu\nu}}{\partial x^\nu} = -4\pi e \left( \overline{\psi}_1 \gamma^\mu \psi_1 - \overline{\psi}_2 \gamma^\mu \psi_2 \right), \tag{3.8}$$

wherefrom the current density is

$$j^\mu = \overline{\psi}_1 \gamma^\mu \psi_1 - \overline{\psi}_2 \gamma^\mu \psi_2\,, \tag{3.9}$$

which conforms with (2.27). The last equation shows that the bispinors $\psi_1$ and $\psi_2$ correspond to charges of opposite signs. At the same time it should be emphasized that the sign in front of $eA_\mu$ is identical in (3.4) and (3.5); the same concerns Eqs. (3.6) and (3.7).

One can deduce from Eqs. (3.4)–(3.7) that

$$\frac{\partial}{\partial x^{\mu}}\left(\overline{\psi}_1\gamma^{\mu}\psi_2+\overline{\psi}_2\gamma^{\mu}\psi_1\right)=0\,. \qquad (3.10)$$

Therefore the condition of (3.2) does not contradict the equations of motion. From (3.4)–(3.7) as well as from (3.8) it follows the equation of continuity $\partial j^{\mu}/\partial x^{\mu} = 0$. Integrating the equation over all space with account taken of the fact that there are no particles at infinity ($\psi_1 = \psi_2 = 0$) we obtain

$$\int\limits_{(\infty)}(\psi_1^*\psi_1-\psi_2^*\psi_2)dV=I\,, \qquad (3.11)$$

where $I$ = constant. Now, if we put $I = 0$, there may be $\psi_1 \neq 0$, $\psi_2 \neq 0$ first and $\psi_1 = 0$, $\psi_2 = 0$ afterwards. Consequently, the equations obtained can depict disappearance of the electronic field, that is to say, the annihilation [cf. Eq. (3.1)].

In this connection it is worthwhile to discuss how the present equations describe a positron. One can verify that, if one introduces bispinors $\psi_1'$ and $\psi_2'$ according to $\psi_1 = C\overline{\psi}_2'$ and $\psi_2 = C\overline{\psi}_1'$, the bispinors $\psi_1'$ and $\psi_2'$ will obey the same equations as above with, however, another sign in front of $A_{\mu}$. Thus, if the above system of equations has a solution $\psi_1$, $\psi_2$, $A_{\mu}$, $v_{\mu}$, the system has the solution $C\overline{\psi}_2$, $C\overline{\psi}_1$, $-A_{\mu}$, $v_{\mu}$ as well. The signs in front of $A_{\mu}$ demonstrate that the solutions are relevant to particles with opposite signs of the charge. Consequently, if the first solution describes the particle (the electron), the second solution describes the antiparticle (the positron). We can assume now that there is an electron in a region of space and a positron in another region, the regions being so far apart that the relevant functions $\psi_1$ and $\psi_2$ do not overlap. It will be shown in the next section that the sign of the constant $I$ in (3.11) depends on the sign of the charge. Let $I = +1$ in the region of the electron and $I = -1$ in the one of the positron. Integrating over all space gives $I = 0$. The subsequent evolution of the system will proceed by the above equations. There may occur the annihilation when $\psi_1$ and $\psi_2$ disappear or a positronium may form when $\psi_1 \neq 0$, $\psi_2 \neq 0$. All of this does not contradict the condition of (3.11). In case we consider only one electron, we must put $I = 1$.

The knowledge of the Lagrangian enables us to find various characteristics of the system under consideration. We begin with the energy-momentum tensor of the system that is defined by [2, 5]

$$T_{\mu}^{\ \nu}=\sum_q q_{,\mu}\frac{\partial L}{\partial q_{,\nu}}-L\delta_{\mu}^{\nu}\,, \qquad (3.12)$$

where $q$ is the set of functions that characterize the system and figure in the Lagrangian $L$ while $q_{,\mu} = \partial q/\partial x^{\mu}$. This formula does not imply the presence of the Lagrange multipliers that enter into $L$ of (3.3). We can however regard the four-vector $v_{\mu}$ not as the set of the Lagrange multipliers but as a function that characterizes the system together with other functions in (3.3). This contradicts nothing for varying $L(x)$ with respect to $v_\mu$ one obtains (3.2). Now the set of the functions $q$ will be $\psi_1$, $\overline{\psi}_1$, $\psi_2$, $\overline{\psi}_2$, $A_\mu$, $v_\mu$. The calculation of $T_\mu^{\ \nu}$ is markedly simplified if account is taken of the fact that substituting (3.4)–(3.7) into (3.3) yields

$$L(x) = -\frac{1}{16\pi} F_{\mu\nu} F^{\mu\nu} . \tag{3.13}$$

Upon raising the lower index with the help of the metric tensor $g^{\mu\nu}$ we obtain, as a result,

$$T^{\mu\nu} = \frac{1}{4\pi}\frac{\partial A_\lambda}{\partial x_\mu} F^{\lambda\nu} + \frac{g^{\mu\nu}}{16\pi} F_{\lambda\eta} F^{\lambda\eta} + \frac{ic\hbar}{2}\left( \overline{\psi}_1 \gamma^\nu \frac{\partial \psi_1}{\partial x_\mu} - \frac{\partial \overline{\psi}_1}{\partial x_\mu}\gamma^\nu \psi_1 - \overline{\psi}_2 \gamma^\nu \frac{\partial \psi_2}{\partial x_\mu} + \frac{\partial \overline{\psi}_2}{\partial x_\mu}\gamma^\nu \psi_2 \right). \tag{3.14}$$

This tensor is neither symmetric nor gauge invariant. To symmetrize it one can add a quantity of the type $\partial\Phi^{\mu\nu\eta}/\partial x^\eta$, where $\Phi^{\mu\nu\eta} = -\,\Phi^{\mu\eta\nu}$ [6]. We take two additional terms

$$\frac{1}{4\pi}\frac{\partial}{\partial x^\eta} A^\mu F^{\nu\eta} = \frac{1}{4\pi}\frac{\partial A^\mu}{\partial x_\eta} F^{\nu}{}_{\eta} - eA^\mu\left(\overline{\psi}_1\gamma^\nu\psi_1 - \overline{\psi}_2\gamma^\nu\psi_2\right), \tag{3.15}$$

$$\frac{ic\hbar}{12}\frac{\partial}{\partial x^\eta}\left[\overline{\psi}_2(\gamma^\mu\sigma^{\nu\eta} + \gamma^\eta\sigma^{\mu\nu} + \gamma^\nu\sigma^{\eta\mu})\psi_2 - \overline{\psi}_1(\gamma^\mu\sigma^{\nu\eta} + \gamma^\eta\sigma^{\mu\nu} + \gamma^\nu\sigma^{\eta\mu})\psi_1\right], \tag{3.16}$$

where

$$\sigma^{\mu\nu} = \frac{1}{2}(\gamma^\mu\gamma^\nu - \gamma^\nu\gamma^\mu) . \tag{3.17}$$

Equation (3.15) has been reduced with regard to (3.8). The idea of the second addition is taken from [7], Sec. 21. The addition is to be transformed with the help of (3.4)–(3.7). Adding the results to (3.14) we obtain finally

$$\begin{aligned} T^{\mu\nu} = {} & \frac{1}{16\pi}\left(g^{\mu\nu}F_{\lambda\eta}F^{\lambda\eta} - 4F^{\mu\lambda}F^{\nu}{}_{\lambda}\right) - \frac{e}{2}\left(A^\mu\overline{\psi}_1\gamma^\nu\psi_1 + A^\nu\overline{\psi}_1\gamma^\mu\psi_1 - A^\mu\overline{\psi}_2\gamma^\nu\psi_2 - A^\nu\overline{\psi}_2\gamma^\mu\psi_2\right) \\ & + \frac{ic\hbar}{4}\left( \overline{\psi}_1\gamma^\nu\frac{\partial\psi_1}{\partial x_\mu} + \overline{\psi}_1\gamma^\mu\frac{\partial\psi_1}{\partial x_\nu} - \frac{\partial\overline{\psi}_1}{\partial x_\mu}\gamma^\nu\psi_1 - \frac{\partial\overline{\psi}_1}{\partial x_\nu}\gamma^\mu\psi_1 - \overline{\psi}_2\gamma^\nu\frac{\partial\psi_2}{\partial x_\mu} - \overline{\psi}_2\gamma^\mu\frac{\partial\psi_2}{\partial x_\nu} \right. \\ & \left. + \frac{\partial\overline{\psi}_2}{\partial x_\mu}\gamma^\nu\psi_2 + \frac{\partial\overline{\psi}_2}{\partial x_\nu}\gamma^\mu\psi_2 \right). \end{aligned} \tag{3.18}$$

This tensor is symmetric and gauge invariant. It enables one to find the density of energy $T^{00}$:

$$T^{00} = \frac{E^2 + H^2}{8\pi} - e\varphi(\psi_1^*\psi_1 - \psi_2^*\psi_2) + \frac{i\hbar}{2}\left( \psi_1^*\frac{\partial\psi_1}{\partial t} - \frac{\partial\psi_1^*}{\partial t}\psi_1 - \psi_2^*\frac{\partial\psi_2}{\partial t} + \frac{\partial\psi_2^*}{\partial t}\psi_2 \right). \tag{3.19}$$

Here we have used the three-dimensional notation: **E** and **H** are the intensities respectively of the electric and magnetic field and φ is the scalar potential.

The symmetricalness of the tensor $T^{\mu\nu}$ allows one to calculate the vector **M** of angular momentum with the help of the formula

$$M_i = \frac{1}{2c}\int\limits_{(\infty)} e_{ijk}(x^j T^{k0} - x^k T^{j0})dV\,. \tag{3.20}$$

Here we have again used the three-dimensional notation: $i, j, k = 1, 2, 3$; $e_{ijk}$ is the antisymmetric unit pseudotensor [6]. Substituting (3.18) here and introducing the vector potential **A** results in

$$\mathbf{M} = \frac{1}{4\pi c}\int\limits_{(\infty)} \big([\mathbf{r}[\mathbf{EH}]] - [\mathbf{rA}]\mathrm{div}\mathbf{E}\big)dV$$

$$+\hbar\int\limits_{(\infty)}\Big\{\tfrac{1}{2}\big(\overline{\psi}_1\gamma^0\boldsymbol{\Sigma}\psi_1 - \overline{\psi}_2\gamma^0\boldsymbol{\Sigma}\psi_2\big) - i\big(\overline{\psi}_1\gamma^0[\mathbf{r}\nabla\psi_1] - \overline{\psi}_2\gamma^0[\mathbf{r}\nabla\psi_2]\big)\Big\}dV\,, \tag{3.21}$$

where as in [2]

$$\boldsymbol{\Sigma} = \begin{pmatrix} \boldsymbol{\sigma} & 0 \\ 0 & \boldsymbol{\sigma} \end{pmatrix}. \tag{3.22}$$

One may check that the vector **M** is gauge invariant.

We now turn to the question as to how the equations should be changed if the system is in an external electromagnetic field. Let the field be created by an external current density $j^{\mu}_{\mathrm{ext}}$. Then $j^{\mu}_{\mathrm{ext}}$ must be added to the current density $j^{\mu}$ that enters into (3.8). We can however proceed in another way. We find $A^{\mathrm{ext}}_{\mu}$ upon solving the equation

$$\frac{\partial F^{\mu\nu}_{\mathrm{ext}}}{\partial x^{\nu}} = -4\pi e j^{\mu}_{\mathrm{ext}}\,. \tag{3.23}$$

Then we can leave Eq. (3.8) unchanged but we must take $A_{\mu} + A^{\mathrm{ext}}_{\mu}$ instead of $A_{\mu}$ in Eqs. (3.4)–(3.7). The second approach is more convenient when the external four-potential $A^{\mathrm{ext}}_{\mu}$ is known in advance.

It will also be of interest in this connection to discuss how one can calculate the energy of the system if the system is in an external field. It is worthwhile to speak of the energy of a system in an external field if the external field does not depend on time. The following considerations show that in the present case one can as before take (3.19) for the density of energy without adding $A^{\mathrm{ext}}_{\mu}$ hereinto. To prove this we take the total energy of the system $E = \int T^{00}dV$ where the integration is carried out over the volume in which the system can be located upon assuming that its motion is finite. Substituting (3.19) therein and making use of the equations of motion (3.4)–(3.7) in which the replacement $A_{\mu} \to A_{\mu} + A^{\mathrm{ext}}_{\mu}$ is made, by tedious but straightforward manipulation one finds that

$$\frac{dE}{dt} + \int \mathrm{div}\mathbf{S}dV = 0, \tag{3.24}$$

where **S** is the Poynting vector:

$$\mathbf{S} = \frac{c}{4\pi}[\mathbf{EH}]. \tag{3.25}$$

The meaning of Eq. (3.24) is simple: the energy of the system remains constant except for a part of the energy that can be carried away by electromagnetic waves [the second term in (3.24)] if they are present. If the external field is time dependent, Eq. (3.24) is to be supplemented with the work of the external forces that depends upon derivatives of $A_\mu^{\mathrm{ext}}$ with respect to time.

## 4. Structure of the electron

According to the argumentation of Introduction, the above equations should, in particular, describe the structure and properties of the electron. To the electron must correspond regular and localized in space (particle-like) solutions to the equations. To the search for such solutions to the coupled Maxwell-Dirac equations as a classical system is devoted rather a vast literature, a thorough review of which can be found in Ref. [8]. However, all particle-like solutions found up till now gave a negative mass $m$ for the relevant formation, which is devoid of physical sense. A solution corresponding to a positive mass is obtained in Ref. [8] but the electric and magnetic fields in the solution are singular as $r \to 0$, which does not allow one to ascribe a physical sense to the solution, let alone the fact that the spherically symmetric solution found in [8] leads to a zero spin of the formation. As mentioned above the most outstanding distinction of the present paper is utilization of two bispinors for description of the electron.

It is sufficient to employ Eqs. (3.4)–(3.5) for the bispinors $\psi_1$ and $\psi_2$ as Eqs. (3.6)–(3.7) follow from the former. It is convenient to work with dimensionless quantities, denoting them by a tilde, according to

$$x^\mu = \lambda\!\!\!^{-}\tilde{x}^\mu,\ \psi_{1,2} = \frac{1}{\lambda\!\!\!^{-3/2}}\tilde{\psi}_{1,2},\ A_\mu = \frac{e}{\lambda\!\!\!^{-}}\tilde{A}_\mu,\ v_\mu = \frac{e}{\lambda\!\!\!^{-}}\tilde{v}_\mu; \quad \mathbf{E} = \frac{e}{\lambda\!\!\!^{-2}}\tilde{\mathbf{E}},\ \mathbf{H} = \frac{e}{\lambda\!\!\!^{-2}}\tilde{\mathbf{H}}, \tag{4.1}$$

where $\lambda\!\!\!^{-} = \hbar/mc$ is the electron Compton wavelength. We recast Eqs. (3.4)–(3.5) in the three-dimensional notation taking into account that $A_\mu = (\varphi, -\mathbf{A})$ and $v_\mu = (v_0, -\mathbf{v})$ [6]:

$$i\frac{\partial\tilde{\psi}_1}{\partial\tilde{t}} + i\boldsymbol{\alpha}\tilde{\nabla}\tilde{\psi}_1 + \alpha\tilde{\mathbf{A}}\boldsymbol{\alpha}\tilde{\psi}_1 - \alpha\tilde{\varphi}\tilde{\psi}_1 - \beta\tilde{\psi}_1 - \alpha\tilde{v}_0\tilde{\psi}_2 + \alpha\tilde{\mathbf{v}}\boldsymbol{\alpha}\tilde{\psi}_2 = 0, \tag{4.2}$$

$$i\frac{\partial\tilde{\psi}_2}{\partial\tilde{t}} + i\boldsymbol{\alpha}\tilde{\nabla}\tilde{\psi}_2 + \alpha\tilde{\mathbf{A}}\boldsymbol{\alpha}\tilde{\psi}_2 - \alpha\tilde{\varphi}\tilde{\psi}_2 - \beta\tilde{\psi}_2 + \alpha\tilde{v}_0\tilde{\psi}_1 - \alpha\tilde{\mathbf{v}}\boldsymbol{\alpha}\tilde{\psi}_1 = 0. \tag{4.3}$$

Here $\alpha = e^2/\hbar c$ is the fine-structure constant while $\boldsymbol{\alpha}$ and $\beta$ are the standard matrices [2]. In what follows we shall omit the tilde bearing in mind that, if an equation does not contain $\hbar$, $c$, $e$ or $m$ but contains $\alpha$, the equation is relevant to the dimensionless quantities.

We shall look for stationary solutions. In this instance one usually assumes that the wave function is proportional to the factor $\exp(-i\varepsilon t/\hbar)$. In our case, however, the whole potential $\varphi$ as well as its constant part is unknown. It follows from (4.2) and (4.3) that the energy $\varepsilon$ can be added to the unknown constant part of $\varphi$ in the fourth term. Keeping this in mind one can merely put $\partial/\partial t = 0$ when seeking the stationary solutions. We seek also axially symmetric solutions (there can be no spherically symmetric solutions for the electron has a spin). An analysis of Eqs. (4.2) and (4.3) shows that, in the cylindrical coordinates $\rho$, $\dot{\varphi}$, $z$, the solution to the equations can be sought in the form (to distinguish between the angle $\varphi$ and the potential $\varphi$ we mark the angle with a dot at the top)

$$\psi_1 = \begin{pmatrix} f_1^{(1)}(\rho, z) \\ f_2^{(1)}(\rho, z) e^{i\dot{\varphi}} \\ i f_3^{(1)}(\rho, z) \\ i f_4^{(1)}(\rho, z) e^{i\dot{\varphi}} \end{pmatrix}, \qquad \psi_2 = \begin{pmatrix} f_1^{(2)}(\rho, z) \\ f_2^{(2)}(\rho, z) e^{i\dot{\varphi}} \\ i f_3^{(2)}(\rho, z) \\ i f_4^{(2)}(\rho, z) e^{i\dot{\varphi}} \end{pmatrix}, \tag{4.4}$$

in which the functions $f_i^{(1,2)}$ are real. Then it follows from (3.9) that $j_\rho = j_z = 0$. There is only one nonzero component of the current density equal to

$$j_{\dot{\varphi}} = 2\left( f_1^{(1)} f_4^{(1)} - f_2^{(1)} f_3^{(1)} - f_1^{(2)} f_4^{(2)} + f_2^{(2)} f_3^{(2)} \right). \tag{4.5}$$

Consequently, the vector potential **A** will have only one nonzero component $A_{\dot{\varphi}}$ as well (denoted henceforth as $A$), which is clear from physical considerations in the event of the axial symmetry. It will be helpful to remark that div **A** = 0 in the present situation. Along similar lines, we shall presume that the vector **v** has only one nonzero component $v_{\dot{\varphi}} = v$, too. Now Eqs. (4.2) and (4.3) in the cylindrical coordinates take the form (we use the standard representation [2] for the matrices)

$$\frac{\partial f_4^{(1)}}{\partial \rho} + \frac{\partial f_3^{(1)}}{\partial z} + \frac{f_4^{(1)}}{\rho} - \alpha A f_4^{(1)} + (1 + \alpha\varphi) f_1^{(1)} + \alpha v_0 f_1^{(2)} - \alpha v f_4^{(2)} = 0, \tag{4.6}$$

$$\frac{\partial f_3^{(1)}}{\partial \rho} - \frac{\partial f_4^{(1)}}{\partial z} + \alpha A f_3^{(1)} + (1 + \alpha\varphi) f_2^{(1)} + \alpha v_0 f_2^{(2)} + \alpha v f_3^{(2)} = 0, \tag{4.7}$$

$$\frac{\partial f_2^{(1)}}{\partial \rho} + \frac{\partial f_1^{(1)}}{\partial z} + \frac{f_2^{(1)}}{\rho} - \alpha A f_2^{(1)} + (1 - \alpha\varphi) f_3^{(1)} - \alpha v_0 f_3^{(2)} - \alpha v f_2^{(2)} = 0, \tag{4.8}$$

$$\frac{\partial f_1^{(1)}}{\partial \rho} - \frac{\partial f_2^{(1)}}{\partial z} + \alpha A f_1^{(1)} + (1 - \alpha\varphi) f_4^{(1)} - \alpha v_0 f_4^{(2)} + \alpha v f_1^{(2)} = 0, \tag{4.9}$$

$$\frac{\partial f_4^{(2)}}{\partial\rho}+\frac{\partial f_3^{(2)}}{\partial z}+\frac{f_4^{(2)}}{\rho}-\alpha A f_4^{(2)}+(1+\alpha\varphi)f_1^{(2)}-\alpha v_0 f_1^{(1)}+\alpha v f_4^{(1)}=0\,, \quad (4.10)$$

$$\frac{\partial f_3^{(2)}}{\partial\rho}-\frac{\partial f_4^{(2)}}{\partial z}+\alpha A f_3^{(2)}+(1+\alpha\varphi)f_2^{(2)}-\alpha v_0 f_2^{(1)}-\alpha v f_3^{(1)}=0\,, \quad (4.11)$$

$$\frac{\partial f_2^{(2)}}{\partial\rho}+\frac{\partial f_1^{(2)}}{\partial z}+\frac{f_2^{(2)}}{\rho}-\alpha A f_2^{(2)}+(1-\alpha\varphi)f_3^{(2)}+\alpha v_0 f_3^{(1)}+\alpha v f_2^{(1)}=0\,, \quad (4.12)$$

$$\frac{\partial f_1^{(2)}}{\partial\rho}-\frac{\partial f_2^{(2)}}{\partial z}+\alpha A f_1^{(2)}+(1-\alpha\varphi)f_4^{(2)}+\alpha v_0 f_4^{(1)}-\alpha v f_1^{(1)}=0\,. \quad (4.13)$$

From (3.8) follow the equations

$$\nabla^2\varphi+4\pi\left(f_1^{(1)2}+f_2^{(1)2}+f_3^{(1)2}+f_4^{(1)2}-f_1^{(2)2}-f_2^{(2)2}-f_3^{(2)2}-f_4^{(2)2}\right)=0\,, \quad (4.14)$$

$$\nabla^2 A-\frac{A}{\rho^2}+8\pi\left(f_1^{(1)}f_4^{(1)}-f_2^{(1)}f_3^{(1)}-f_1^{(2)}f_4^{(2)}+f_2^{(2)}f_3^{(2)}\right)=0\,. \quad (4.15)$$

The conditions of (3.2) furnish only two equations

$$f_1^{(1)}f_1^{(2)}+f_2^{(1)}f_2^{(2)}+f_3^{(1)}f_3^{(2)}+f_4^{(1)}f_4^{(2)}=0\,, \quad (4.16)$$

$$f_1^{(1)}f_4^{(2)}-f_2^{(1)}f_3^{(2)}-f_3^{(1)}f_2^{(2)}+f_4^{(1)}f_1^{(2)}=0\,. \quad (4.17)$$

These two equations suffice insofar as the four-vector $v^\mu$ has now only two nonzero components $v_0$ and $v$. For this reason, the number of unknowns (there are 12 unknowns) is equal to the number of Eqs. (4.6)–(4.17). The equations must be supplemented with the normalization condition that follows from (3.11):

$$\int\limits_{(\infty)}\left(f_1^{(1)2}+f_2^{(1)2}+f_3^{(1)2}+f_4^{(1)2}-f_1^{(2)2}-f_2^{(2)2}-f_3^{(2)2}-f_4^{(2)2}\right)dV=1\,. \quad (4.18)$$

Hereinafter the integral on the left will be denoted as $I$.

First of all, it is of interest to discuss properties of the formation described by a particle-like solution to the equations obtained if such a solution exists. It is worthy of remark that this solution implies that the functions $f_i^{(1,2)}$ decrease sufficiently rapidly with distance $r$ from the center in order that the integral $I$ in (4.18) exist (as will be shown below the functions decrease exponentially with increasing $r$). We presume also that none of the occurring functions have singularities anywhere the point $r = 0$ inclusive.

We rewrite Eq. (4.14) in spherical coordinates $r$, $\theta$, $\dot{\varphi}$ with account taken of the fact that the potential $\varphi$ does not depend on the angle $\dot{\varphi}$:

$$\frac{1}{r^2}\frac{\partial}{\partial r}\left(r^2\frac{\partial\varphi}{\partial r}\right)+\frac{1}{r^2\sin\theta}\frac{\partial}{\partial\theta}\left(\sin\theta\frac{\partial\varphi}{\partial\theta}\right)$$

$$+4\pi\left(f_1^{(1)^2}+f_2^{(1)^2}+f_3^{(1)^2}+f_4^{(1)^2}-f_1^{(2)^2}-f_2^{(2)^2}-f_3^{(2)^2}-f_4^{(2)^2}\right)=0. \qquad (4.19)$$

We multiply the equation by the volume element in the spherical coordinates $dV$ = $r^2\sin\theta dr d\theta d\dot{\varphi}$ and integrate over all angles and over $r$ from 0 to such $r$ where all the functions $f_i^{(1,2)}$ practically vanish. Integration of the second term in (4.19) over θ from 0 to π gives zero at once while we can integrate the first term with respect to $r$. The potential φ must not depend on the angle θ as $r \to \infty$, which leads to the equation

$$r^2\frac{\partial\varphi}{\partial r}+I=0. \qquad (4.20)$$

Integrating this equation yields $\varphi = I/r$ + constant, or $\varphi = e/r$ + constant in the dimensional units where the normalization condition of (4.18) is taken into account. The last asymptotic expression indicates that the formation in question has the same charge $e$ as the electron whereas the same quantity $e$ enters into the starting Eqs. (3.4)–(3.8).

In order to find a formula for the magnetic moment $\mu$ we remark that our definition of the current density $j^\mu$ of (3.9) and the definition used in [6] and denoted as $j_{\rm L}^\mu$ are interrelated by $j_{\rm L}^\mu = ecj^\mu$. Now referring to [6] we have $\boldsymbol{\mu} = e\lambda\tilde{\boldsymbol{\mu}}$ with the dimensionless quantity

$$\tilde{\boldsymbol{\mu}}=\tfrac{1}{2}\int\limits_{(\infty)}[\mathbf{rj}]dV. \qquad (4.21)$$

With regard to the fact that the current density has only one component (4.5), upon integrating over the angle $\dot{\varphi}$ one finds that $\tilde{\boldsymbol{\mu}}$ has only one component as well which is of the form

$$\tilde{\mu}_z=\int\limits_{(\infty)}\rho\left(f_1^{(1)}f_4^{(1)}-f_2^{(1)}f_3^{(1)}-f_1^{(2)}f_4^{(2)}+f_2^{(2)}f_3^{(2)}\right)dV. \qquad (4.22)$$

It is worth remarking that the same result can also be obtained from (4.15) by taking into consideration that the vector potential is of the form $\mathbf{A} = [\boldsymbol{\mu}\mathbf{r}]/r^3$ as $r \to \infty$ [6]. To calculate $\mu_z$ it is necessary to solve the above equations. An analogical situation occurs in QED where the electron magnetic moment with account taken of its anomalous part is calculated upon solving the QED equations.

We turn now to the angular momentum of the formation. It may be shown that the first integral in (3.21) vanishes for the stationary solution under consideration with account taken of the fact that φ and **A** behave at infinity as pointed out above. Upon introducing (4.4) into the second integral and integrating over the angle $\dot{\varphi}$ we obtain that there is only one nonzero component of the vector **M** equal to $M_z = \hbar I/2$ in the ordinary units. By (4.18) we see that the angular momentum of the formation is the same as the one of the electron $\hbar/2$. Thus the charge and spin of the formation coincide with those of the electron, although the bispinors $\psi_1$ and $\psi_2$ correspond to charges of different signs.

Equations (4.2)–(4.3) at $\alpha = 0$ reduce to Dirac's equations for a free particle [see also Eqs. (3.4)–(3.5) at $e = 0$]. The Dirac equation for free particles has no particle-like solutions. Therefore, if Eqs. (4.2)–(4.3) have a particle-like solution, it should come into existence at some finite $\alpha$. It will be very strange if this value of $\alpha$ differs from the well-known value $\alpha \approx 1/137$. If the particle-like solution to the above equations is discovered at $\alpha \approx 1/137$ (the solution can naturally be found only numerically), this will explain the value of $\alpha$ as a value at which the free electron exists. Up to the present no convincing argumentation was known as to why the fundamental physical constant $\alpha$ has just this magnitude $\approx 1/137$.

In this connection it is worthwhile to point out a property of the above system of equations. Let the system admit a particle-like solution $f_i^{(1,2)}$, $\varphi$, $A$, $v_0$, $v$ at a definite value of $\alpha$. Then the system will have such a solution at another $\alpha'$ with

$$\alpha' = \frac{\alpha}{C}, \quad f_i'^{(1,2)} = \sqrt{C} f_i^{(1,2)}, \quad \varphi' = C\varphi, \quad A' = CA, \quad \nu_0' = C\nu_0, \quad v' = Cv\,, \tag{4.23}$$

wherein $C$ is an arbitrary number. The last solution, however, does not make direct physical sense inasmuch as $I' = CI$ in this event while, in order that the charge and spin of the formation coincide with those of the electron, it is necessary that $I' = 1$ as in (4.18). At the same time this property may prove to be helpful if one solves the equations numerically. The point is that when solving the equations numerically it is unclear how to find the required value of $\alpha$, and, besides, an additional difficulty consists in the necessity of choosing, in the course of the solution, functions that would satisfy the normalization condition of (4.18). One can circumvent these difficulties simultaneously by putting, for example, $C = \alpha$ in (4.23) when $\alpha' = 1$. This means that when solving the equations numerically one can set $\alpha = 1$ in them (which, by the way, simplifies formulae) without taking care to satisfy the normalization condition of (4.18). Upon finding the particle-like solution one calculates the integral $I'$ [that is, the integral that figures on the left side of (4.18)] thus obtaining immediately the sought value of $\alpha$ because $I' = \alpha$ now. The actual magnitude of the functions may be computed by a transformation inverse with respect to (4.23).

We now proceed to a close examination of the above equations. Equations (4.6) and (4.8) contain the terms $f_4^{(1)}/\rho$ and $f_2^{(1)}/\rho$ that will not be singular as $\rho \to 0$ only if $f_4^{(1)} \propto \rho$ and $f_2^{(1)} \propto \rho$. A further investigation suggests the following structure of the functions

$$f_1^{(1)} = g_1^{(1)}(\rho^2, z^2),\; f_2^{(1)} = \rho z g_2^{(1)}(\rho^2, z^2),\; f_3^{(1)} = z g_3^{(1)}(\rho^2, z^2),\; f_4^{(1)} = \rho g_4^{(1)}(\rho^2, z^2)\,,$$

$$\varphi = \varphi(\rho^2, z^2),\; A = \rho\gamma(\rho^2, z^2)\,. \tag{4.24}$$

The structure of $A$ results from the asymptotic formula pointed out after (4.22). As to the structure of the functions $f_i^{(2)}$, the uniqueness is absent here even if the formulae of (4.24) are taken into account although it is clear from (4.10) and (4.12) that $f_4^{(2)} \propto \rho$ and $f_2^{(2)} \propto \rho$. An extra help is provided by Eqs. (4.16) and (4.17). The key idea was based on the assumption that, in the ground state, the functions must contain the least possible number of the factors ρ and $z$. As a result, we are led to

$$f_1^{(2)} = zg_1^{(2)}(\rho^2, z^2),\ f_2^{(2)} = \rho g_2^{(2)}(\rho^2, z^2),\ f_3^{(2)} = g_3^{(2)}(\rho^2, z^2),\ f_4^{(2)} = \rho z g_4^{(2)}(\rho^2, z^2),$$

$$v_0 = zu_0(\rho^2, z^2),\ v = \rho z u(\rho^2, z^2). \tag{4.25}$$

The structure of $v_0$ and $v$ follows now directly from the structure of the other functions.

As has already been intimated we imply that the energy ε that figures in the factor $\exp(-i\varepsilon t/\hbar)$ for stationary solutions is added to the constant part of the potential φ. But the relativistic energy contains the rest energy $mc^2$ as well. It is desirable to single out the last energy explicitly in φ. To this end we remark that the factor $\exp(-i\varepsilon t/\hbar)$ with $\varepsilon = mc^2$ becomes $\exp(-i\tilde{t})$ in the dimensionless units of (4.1). Upon placing the last factor in the first term in (4.2) and (4.3) we obtain 1 in the front of $\exp(-i\tilde{t})$. Therefore the summand in $-\alpha\tilde{\varphi}$ of (4.2) and (4.3) that is due to the rest energy is equal to 1, which amounts to saying that the potential is of the form

$$\tilde{\varphi} = -\frac{1}{\alpha} + c_0 + \overline{\varphi}(\mathbf{r}), \tag{4.26}$$

wherein $c_0$ is a comparatively small constant and $\overline{\varphi}(\mathbf{r}) \to 0$ as $r \to \infty$.

In support of the reasonableness of that expression we find the energy of the formation $\mathcal{E} = \int T^{00} dV$ with use made of (3.19) where $\partial/\partial t = 0$ and with account taken of (4.26) and (3.11) where $I = 1$ [see also (4.18)]. In the ordinary units

$$\mathcal{E} = \frac{c\hbar}{\lambda\!\!\!^{-}}\left\{1 - \alpha c_0 + \alpha \int\limits_{(\infty)} \left[\frac{1}{8\pi}\left(\tilde{E}^2 + \tilde{H}^2\right) - \overline{\varphi}\left(\tilde{\psi}_1^*\tilde{\psi}_1 - \tilde{\psi}_2^*\tilde{\psi}_2\right)\right] d\tilde{V}\right\}. \tag{4.27}$$

For clarity sake, here again we have utilized the tilde to denote the dimensionless quantities of (4.1). If $\alpha = 0$, we obtain that $\mathcal{E} = mc^2$ as $\lambda\!\!\!^{-} = \hbar/mc$. Equation (4.27) will be further discussed in Sec. 5.

We can now write down the equations for the new functions that follow from (4.6)–(4.13):

$$\rho\frac{\partial g_4^{(1)}}{\partial\rho} + z\frac{\partial g_3^{(1)}}{\partial z} + 2g_4^{(1)} + g_3^{(1)} - \alpha\rho^2\gamma g_4^{(1)} + \alpha\left(c_0 + \overline{\varphi}\right)g_1^{(1)} + \alpha z^2 u_0 g_1^{(2)} - \alpha\rho^2 z^2 u g_4^{(2)} = 0, \tag{4.28}$$

$$\frac{\partial g_3^{(1)}}{\rho\partial\rho}-\frac{\partial g_4^{(1)}}{z\partial z}+\alpha\gamma g_3^{(1)}+\alpha\left(c_0+\overline{\varphi}\right)g_2^{(1)}+\alpha u_0 g_2^{(2)}+\alpha u g_3^{(2)}=0\,, \tag{4.29}$$

$$\rho\frac{\partial g_2^{(1)}}{\partial\rho}+\frac{\partial g_1^{(1)}}{z\partial z}+2g_2^{(1)}-\alpha\rho^2\gamma g_2^{(1)}+\left(2-\alpha c_0-\alpha\overline{\varphi}\right)g_3^{(1)}-\alpha u_0 g_3^{(2)}-\alpha\rho^2 u g_2^{(2)}=0\,, \tag{4.30}$$

$$\frac{\partial g_1^{(1)}}{\rho\partial\rho}-z\frac{\partial g_2^{(1)}}{\partial z}-g_2^{(1)}+\alpha\gamma g_1^{(1)}+\left(2-\alpha c_0-\alpha\overline{\varphi}\right)g_4^{(1)}-\alpha z^2 u_0 g_4^{(2)}+\alpha z^2 u g_1^{(2)}=0\,, \tag{4.31}$$

$$\rho\frac{\partial g_4^{(2)}}{\partial\rho}+\frac{\partial g_3^{(2)}}{z\partial z}+2g_4^{(2)}-\alpha\rho^2\gamma g_4^{(2)}+\alpha\left(c_0+\overline{\varphi}\right)g_1^{(2)}-\alpha u_0 g_1^{(1)}+\alpha\rho^2 u g_4^{(1)}=0\,, \tag{4.32}$$

$$\frac{\partial g_3^{(2)}}{\rho\partial\rho}-z\frac{\partial g_4^{(2)}}{\partial z}-g_4^{(2)}+\alpha\gamma g_3^{(2)}+\alpha\left(c_0+\overline{\varphi}\right)g_2^{(2)}-\alpha z^2 u_0 g_2^{(1)}-\alpha z^2 u g_3^{(1)}=0\,, \tag{4.33}$$

$$\rho\frac{\partial g_2^{(2)}}{\partial\rho}+z\frac{\partial g_1^{(2)}}{\partial z}+g_1^{(2)}+2g_2^{(2)}-\alpha\rho^2\gamma g_2^{(2)}+\left(2-\alpha c_0-\alpha\overline{\varphi}\right)g_3^{(2)}+\alpha z^2 u_0 g_3^{(1)}+\alpha\rho^2 z^2 u g_2^{(1)}=0\,, \tag{4.34}$$

$$\frac{\partial g_1^{(2)}}{\rho\partial\rho}-\frac{\partial g_2^{(2)}}{z\partial z}+\alpha\gamma g_1^{(2)}+\left(2-\alpha c_0-\alpha\overline{\varphi}\right)g_4^{(2)}+\alpha u_0 g_4^{(1)}-\alpha u g_1^{(1)}=0\,. \tag{4.35}$$

Equations (4.14) and (4.15) reduce to

$$\nabla^2\overline{\varphi}+4\pi\left(g_1^{(1)2}+\rho^2 z^2 g_2^{(1)2}+z^2 g_3^{(1)2}+\rho^2 g_4^{(1)2}-z^2 g_1^{(2)2}-\rho^2 g_2^{(2)2}-g_3^{(2)2}-\rho^2 z^2 g_4^{(2)2}\right)=0\,, \tag{4.36}$$

$$\frac{\partial^2\gamma}{\partial\rho^2}+\frac{3}{\rho}\frac{\partial\gamma}{\partial\rho}+\frac{\partial^2\gamma}{\partial z^2}+8\pi\left(g_1^{(1)}g_4^{(1)}-z^2 g_2^{(1)}g_3^{(1)}-z^2 g_1^{(2)}g_4^{(2)}+g_2^{(2)}g_3^{(2)}\right)=0\,. \tag{4.37}$$

It follows from (4.16) and (4.17) that

$$g_1^{(2)}=-\rho^2 B_1 g_2^{(2)}-B_2 g_3^{(2)},\quad g_4^{(2)}=B_2 g_2^{(2)}+B_1 g_3^{(2)}\,, \tag{4.38}$$

where

$$B_1=\frac{g_1^{(1)}g_2^{(1)}+g_3^{(1)}g_4^{(1)}}{D},\quad B_2=\frac{g_1^{(1)}g_3^{(1)}+\rho^2 g_2^{(1)}g_4^{(1)}}{D},\quad D=g_1^{(1)2}-\rho^2 g_4^{(1)2}\,. \tag{4.39}$$

These expressions for $g_1^{(2)}$ and $g_4^{(2)}$ can be introduced into all other equations so that the number of the unknown functions is reduced to 10.

Solving Eqs. (4.32) and (4.35) simultaneously for $u_0$ and $u$ we obtain

$$\alpha u_0=\frac{U_1 g_1^{(1)}-\rho^2 U_2 g_4^{(1)}}{D},\quad \alpha u=\frac{U_1 g_4^{(1)}-U_2 g_1^{(1)}}{D}\,, \tag{4.40}$$

where

$$U_1=\rho B_2\frac{\partial g_2^{(2)}}{\partial\rho}+\rho B_1\frac{\partial g_3^{(2)}}{\partial\rho}+\frac{\partial g_3^{(2)}}{z\partial z}+\left[2B_2+\rho\frac{\partial B_2}{\partial\rho}-\alpha\rho^2\gamma B_2-\alpha\rho^2(c_0+\overline{\varphi})B_1\right]g_2^{(2)}$$

$$+\left[2B_1+\rho\frac{\partial B_1}{\partial\rho}-\alpha\rho^2\gamma B_1-\alpha(c_0+\overline{\varphi})B_2\right]g_3^{(2)}, \tag{4.41}$$

$$U_2=\rho B_1\frac{\partial g_2^{(2)}}{\partial\rho}+\frac{\partial g_2^{(2)}}{z\partial z}+B_2\frac{\partial g_3^{(2)}}{\rho\partial\rho}+\left[2B_1+\rho\frac{\partial B_1}{\partial\rho}+\alpha\rho^2\gamma B_1-(2-\alpha c_0-\alpha\overline{\varphi})B_2\right]g_2^{(2)}$$

$$+\left[\frac{\partial B_2}{\rho\partial\rho}+\alpha\gamma B_2-(2-\alpha c_0-\alpha\overline{\varphi})B_1\right]g_3^{(2)}. \tag{4.42}$$

If these $u_0$ and $u$ are substituted into all other equations the number of the unknown functions will be reduced to 8. In particular, if the substitution is made into Eqs. (4.33) and (4.34), a rather lengthy calculation will lead to the equations

$$\rho C\frac{\partial g_2^{(2)}}{\partial\rho}-2\rho^2 zB_1\frac{\partial g_2^{(2)}}{\partial z}+C_{11}g_2^{(2)}+C_{12}g_3^{(2)}=0, \tag{4.43}$$

$$C\frac{\partial g_3^{(2)}}{\rho\partial\rho}-2zB_1\frac{\partial g_3^{(2)}}{\partial z}+C_{21}g_2^{(2)}+C_{22}g_3^{(2)}=0, \tag{4.44}$$

in which

$$C=1+z^2B_3,\quad B_3=\frac{g_3^{(1)^2}-\rho^2 g_2^{(1)^2}}{D}, \tag{4.45}$$

$$C_{11}=2+\rho z^2\left(B_2\frac{\partial B_2}{\partial\rho}-\rho^2B_1\frac{\partial B_1}{\partial\rho}\right)-\rho^2 z\frac{\partial B_1}{\partial z}-\rho^2B_1+2z^2B_3-\alpha\rho^2\gamma(1+\rho^2z^2B_1^2+z^2B_2^2)$$

$$+2\rho^2z^2(1-\alpha c_0-\alpha\overline{\varphi})B_1B_2, \tag{4.46}$$

$$C_{12}=\rho z^2\left(B_2\frac{\partial B_1}{\partial\rho}-B_1\frac{\partial B_2}{\partial\rho}\right)-z\frac{\partial B_2}{\partial z}-B_2+2z^2B_1B_2(1-\alpha\rho^2\gamma)$$

$$+2(1+\rho^2z^2B_1^2)-\alpha(c_0+\overline{\varphi})(1+\rho^2z^2B_1^2+z^2B_2^2), \tag{4.47}$$

$$C_{21}=\rho z^2\left(B_2\frac{\partial B_1}{\partial\rho}-B_1\frac{\partial B_2}{\partial\rho}\right)-z\frac{\partial B_2}{\partial z}-B_2+2\alpha\rho^2z^2\gamma B_1B_2$$

$$-2z^2B_2^2+\alpha(c_0+\overline{\varphi})(1+\rho^2z^2B_1^2+z^2B_2^2), \tag{4.48}$$

$$C_{22}=z^2\left(B_2\frac{\partial B_2}{\rho\partial\rho}-\rho B_1\frac{\partial B_1}{\partial\rho}\right)-z\frac{\partial B_1}{\partial z}-B_1-2z^2B_1^2+\alpha\gamma(1+\rho^2z^2B_1^2+z^2B_2^2)$$

$$-2z^2(1-\alpha c_0-\alpha\overline{\varphi})B_1B_2. \tag{4.49}$$

Although Eqs. (4.43) and (4.44) look like linear equations for $g_2^{(2)}$ and $g_3^{(2)}$, in actual fact they are nonlinear equations because the quantities $\overline{\varphi}$ and $\gamma$ that enter into the coefficients (4.46)–(4.49) depend upon $g_2^{(2)}$ and $g_3^{(2)}$. From these equations one can deduce second-order partial differential equations separately for $g_2^{(2)}$ and $g_3^{(2)}$. The equations obtained turn out to be differential equations of parabolic type whereas all other equations bear the elliptic character (see below).

Equations (4.36) and (4.37) can be solved for $\overline{\varphi}$ and γ since such equations are well-known from the theory of electromagnetism. Equation (4.36) is of the form of the Poisson equation and, when written as $\nabla^2\overline{\varphi} + 4\pi Q = 0$, it has the solution [6]

$$\overline{\varphi} = \int\limits_{(\infty)} \frac{Q(\mathbf{r}')}{|\mathbf{r}-\mathbf{r}'|} d\mathbf{r}' . \quad (4.50)$$

It should be underlined that this formula yields the necessary for us solution that vanishes at infinity. As to γ, it is convenient to start from the vectorial equation $\nabla^2\mathbf{A} + 4\pi\mathbf{j} = 0$ of which each component satisfies an equation of the type (4.50). With due regard for the fact that the vector **j** in the present case has only one component (4.5) one obtains

$$A_{\dot{\varphi}} \equiv A = \int\limits_{(\infty)} \frac{\cos(\dot{\varphi}-\dot{\varphi}') j_{\dot{\varphi}}(\mathbf{r}')}{|\mathbf{r}-\mathbf{r}'|} d\mathbf{r}' . \quad (4.51)$$

It may be shown that this integral is proportional to ρ, which enables one to find γ because $A = \rho\gamma$. In concrete calculations, it is convenient to expand the quantity $1/|\mathbf{r} - \mathbf{r}'|$ in (4.50) and (4.51) in terms of spherical harmonics (see, e.g., § 41 in [6]).

If $\overline{\varphi}$ and γ as given by (4.50) and (4.51) are placed in the remaining equations, we shall obtain six equations (4.28)–(4.31), (4.43), (4.44) for six functions $g_1^{(1)}, g_2^{(1)}, g_3^{(1)}, g_4^{(1)}, g_2^{(2)}, g_3^{(2)}$. It should be observed that, even if we set $\alpha' = 1$ (see above), in the equations there remains the unknown constant $c_0$ from (4.26) which should be found in the course of solving the equations with the proviso that the occurring functions are nonsingular and vanish sufficiently rapidly at infinity. Although the above equations are written down in the cylindrical system of coordinates, it is convenient sometimes to exploit the spherical coordinates $r, \theta, \dot{\varphi}$ in terms of which the equations can be readily recast.

From Eqs. (4.28)–(4.31) one can deduce second-order differential equations separately for the functions $g_i^{(1)}$ each. The simplest way to do this is to proceed from (4.6)–(4.9). Say, for $f_1^{(1)}$ we obtain

$$\nabla^2 f_1^{(1)} - \nu^2 f_1^{(1)} + Q_1 = 0 , \quad (4.52)$$

where

$$\nu^2 = \alpha c_0 (2 - \alpha c_0) , \quad (4.53)$$

$$Q_1 = -\rho \frac{\partial G_4^{(1)}}{\partial \rho} - z \frac{\partial G_3^{(1)}}{\partial z} - 2G_4^{(1)} - G_3^{(1)} - (2 - \alpha c_0) G_1^{(1)} . \quad (4.54)$$

Here, for brevity sake, the following notation is introduced

$$G_1^{(1)} = \alpha\overline{\varphi}g_1^{(1)} - \alpha\rho^2\gamma g_4^{(1)} + \alpha z^2 u_0 g_1^{(2)} - \alpha\rho^2 z^2 u g_4^{(2)}, \tag{4.55}$$

$$G_2^{(1)} = \alpha\overline{\varphi}g_2^{(1)} + \alpha\gamma g_3^{(1)} + \alpha u_0 g_2^{(2)} + \alpha u g_3^{(2)}, \tag{4.56}$$

$$G_3^{(1)} = \alpha\overline{\varphi}g_3^{(1)} + \alpha\rho^2\gamma g_2^{(1)} + \alpha u_0 g_3^{(2)} + \alpha\rho^2 u g_2^{(2)}, \tag{4.57}$$

$$G_4^{(1)} = \alpha\overline{\varphi}g_4^{(1)} - \alpha\gamma g_1^{(1)} + \alpha z^2 u_0 g_4^{(2)} - \alpha z^2 u g_1^{(2)}. \tag{4.58}$$

The remaining equations are written down in Appendix A.

With use made of the relevant Green function Eq. (4.52) can be rewritten in the integral form:

$$f_1^{(1)} = \frac{1}{4\pi}\int\limits_{(\infty)} \frac{e^{-\nu|\mathbf{r}-\mathbf{r}'|}}{|\mathbf{r}-\mathbf{r}'|} Q_1(\mathbf{r}')d\mathbf{r}'. \tag{4.59}$$

The efficiency of such a representation when solving an equation of the type (4.52) simultaneously with an equation of the type (4.50) was demonstrated in Ref. [9]. The method of iterations was exploited which permitted one to calculate concurrently the constant $\nu$ connected with our unknown constant $c_0$ by (4.53). At the same time, the situation in the present paper is, of course, much more complicated inasmuch as we have to solve four equations of the type (4.59) rather than one, namely, Eqs. (4.59) and (A.5).

We turn now to Eqs. (4.43) and (4.44). The partial derivatives with respect to $z$ disappear when $z = 0$ so that the equations reduce to a system of ordinary differential equations:

$$\rho\frac{dg_2^{(2)}}{d\rho} + C_{11}g_2^{(2)} + C_{12}g_3^{(2)} = 0, \tag{4.60}$$

$$\frac{dg_3^{(2)}}{\rho d\rho} + C_{21}g_2^{(2)} + C_{22}g_3^{(2)} = 0, \tag{4.61}$$

in which the coefficients must be taken at $z = 0$ where they acquire rather a simple form. This system of linear equations can be readily solved by standard numerical methods, for example, by Runge-Kutta methods. Choosing the test functions $g_1^{(1)}, g_2^{(1)}, g_3^{(1)}, g_4^{(1)}, g_2^{(2)}, g_3^{(2)}$ in a reasonable form that can be found by analyzing the above equations in the limit as $r \to 0$ and $r \to \infty$ and computing the coefficients in (4.60) and (4.61) as functions of $\rho = r$, we shall see that the solutions to the equations generally diverge as $\rho \to \infty$. But upon changing one of the parameters, for instance upon multiplying one of the test functions by a constant, we can obtain converging solutions although this does not always happen. In particular, by the constant can be multiplied the test functions $g_2^{(2)}$ and $g_3^{(2)}$ simultaneously. It would seem that this will not change Eqs.

(4.60)–(4.61) but this will change $\overline{\varphi}$ and $\gamma$ in (4.50) and (4.51) and thereby the coefficients in (4.60)–(4.61).

As mentioned above Eqs. (4.43)–(4.44) pertain to differential equations of parabolic type. In the present case the characteristic equation is of the form

$$\frac{dz}{d\rho} = -\frac{2\rho z B_1}{C}. \tag{4.62}$$

If one moves along a characteristic, Eqs. (4.43)–(4.44) reduce again to a system of ordinary differential equations of the type (4.60)–(4.61) (the factor $C$ will appear in front of the derivatives), which is possible just for partial differential equations of the parabolic type. One can begin the characteristics on the $z$-axis with different $z$, and it is convenient to so choose the test functions that $C \neq 0$, in which case the characteristics will be smooth curves. In this case too, upon multiplying the test functions $g_2^{(2)}$ and $g_3^{(2)}$ simultaneously by a constant factor it is possible to attain that the solutions to the equations in question vanish as $\rho \to \infty$. However, the factors at $z = 0$ and at initial $z \neq 0$ turn out to be different. Hence the main difficulty is to find solutions to Eqs. (4.43)–(4.44) such that they vanish at infinity in overall space, and this must be done simultaneously with solving the equations for $g_i^{(1)}$.

It is seen from Eqs. (4.59) and (A.5) that the behavior of the functions as $r \to \infty$ is determined first of all by the factor $e^{-\nu r}$. It may be shown as well that localized solutions to Eqs. (4.60) and (4.61) are proportional to $e^{-\nu\rho}$ as $\rho \to \infty$. Therefore the size of the particle-like solution that should represent the electron is in magnitude of the order $1/\nu$ in the dimensionless units. It follows from (4.53) that always $\nu \leq 1$and $\nu = 1$ if $c_0 = 1/\alpha$, in which case the second term in (4.26) equals the first in magnitude. As mentioned in connection with (4.26) the constant $c_0$ should be comparatively small. If we take $c_0 \approx 1$, there will result $1/\nu \approx 8$. Preliminary numerical calculations demonstrate that $c_0$ may be of the order 0.2, then $1/\nu$ will be about 20. According to (4.1) the unit of length in our case is the electron Compton wavelength $\lambda\!\!\!^{-}$. It should be remarked for comparison that the Bohr radius in a hydrogen atom is $\hbar^2/me^2 = \lambda\!\!\!^{-}/\alpha \approx 137\lambda\!\!\!^{-}$. Hence, in any case the size of the free electron lies in the range between the electron Compton wavelength and the Bohr radius.

Such a large size of the electron may cause surprise since it is believed thus far that the electron size is so small that various attempts to find it have not met with success. To estimate the electron size one utilizes sometimes the classical radius of the electron $e^2/mc^2 = \alpha\lambda\!\!\!^{-} \approx 3{\cdot}10^{-13}$ cm [6]. Let us discuss, however, how one can determine the size of a particle. This size can be found out, for example, when the particle collides with other particles. One may mention, as a well-known example, the determination of the size of atomic nuclei by Rutherford in 1913. We must, in our case, establish the laws of the particle-by-particle scattering with used made of quantum mechanical formulae under the assumption that electromagnetic forces alone act between the particles. One should, in actual fact, resort to QED equations in which only electromagnetic forces are taken into account, or to the equations obtained in the present paper. If deviations from the laws established are observed at some impact parameters, one can assert that one has, at these impact parameters, penetrated into the particle itself where other forces act. From such experiments, one can find out the size of the particle. It should be stressed that creation of new particles is possible at high energies but such events must be rejected as long as this is beyond the scope of QED and of our equations. Up till now no deviations from formulae calculated according to the recipes of QED were reported in experiments with electrons. Moreover, there is a remarkable accord between the QED calculations and experiment, the accord that has no analogues with other theories. All of this points out seemingly that the size of the electron should be extremely small. In particular, there are experiments indicating that the electron radius is less than $6 \cdot 10^{-22}$ cm [10].

In this connection it is reasonable to assume that the electron size is equal to zero altogether, which amounts to saying that the electron has no foreign core. As long as the electron does exist nevertheless, the only possibility that remains is that the structure of the electron is wholly described by the above functions $f_i^{(1,2)}$; in particular, the density of the electronic substance is equal to the integrand in (4.18). Upon arriving at this conclusion it is not difficult to ascertain that not only does the size of the electron found above not contradict experiment, but on the contrary it is supported by all experimental data concerning quantum mechanics. Carrying out calculation of an effect according to equations of quantum mechanics (or QED) we obtain a result in complete agreement with experiment. In the course of the calculation, however, we have explicitly or implicitly used the wave function whereas just this function determines the electron size, which amounts to saying that the electron size was taken into account. In other words, the electron size manifests itself in all quantum mechanical experiments with electrons. A free electron has the size lying in the range between the electron Compton wavelength and the Bohr radius, and resembles a cloud. When meeting with a proton the electronic cloud envelops it, thus giving rise to a hydrogen atom. Curiously, the sizes of the electronic cloud increase in this event (up to the Bohr radius) in spite of the attraction to the proton. As to the question how such a picture conforms with the conventional probabilistic interpretation of the wave function, the author hopes to consider it elsewhere.

It is of interest to discuss the question as to why the electron remains localized in space despite the electric repulsion between its parts charged similarly. In the electron there exists an electric current whose density is equal to $ej_\varphi$ with $j_\varphi$ given by (4.5). At the same time, currents flowing in the same direction attract. The attractive magnetic force arising between moving like charges is, however, proportional to $v^2/c^2$ where $v$ is the velocity of the charges [6] and thereby it is generally much weaker than the repulsive electric force between the charges. On the other hand, if one calculates the velocity of a particle with use made of the Dirac equation, the velocity will always appear to be equal to the velocity of light $c$ [11]. The really observable velocity will be less than $c$, of course, but one must formally introduce $v = c$ into equations. As a consequence, the magnetic attraction of the currents in the electron will be of the same order of magnitude as the electric repulsion and can compensate the later. The magnetic attraction alone is, however, insufficient because on the rotation axis, for example, the current vanishes [ $j_\varphi \propto \rho$ according to (4.5)] and the magnetic attraction is absent. Consequently, an amount of charge of opposite sign is required which is provided by the second bispinor $\psi_2$ that the charge of another sign corresponds to. This argumentation indicates an added physical reason as to why the electron must be described by two bispinors $\psi_1$ and $\psi_2$ besides that considered at the outset of Sec. 3.

We return to the positron once more. As was pointed out in Sec. 3 when discussing Eq. (3.11) the positron is described by bispinors $\psi_1'$ and $\psi_2'$ defined by the relations $\psi_1 = C\overline{\psi}_2'$ and $\psi_2 = C\overline{\psi}_1'$ while the sign in front of $A_\mu$ should be changed in equations (the sign in front of $v_\mu$ remains the same). In particular, Eq. (3.8) acquires the form

$$\frac{\partial F^{\mu\nu}}{\partial x^\nu} = -4\pi e\left(\overline{\psi}_2'\gamma^\mu\psi_2' - \overline{\psi}_1'\gamma^\mu\psi_1'\right). \tag{4.63}$$

It follows therefrom a condition of the type (3.11) with $I = -1$ used when discussing (3.11). We introduce dimensionless quantities as in (4.1) but with another sign in $A_\mu$:

$$x^\mu = \lambda\!\!\!^{-}\tilde{x}^\mu,\ \psi_{1,2}' = \frac{1}{\lambda\!\!\!^{-\,3/2}}\tilde{\psi}_{1,2},\ A_\mu = -\frac{e}{\lambda\!\!\!^{-}}\tilde{A}_\mu,\ v_\mu = \frac{e}{\lambda\!\!\!^{-}}\tilde{v}_\mu\,. \tag{4.64}$$

As a result we obtain equations exactly identical with the above equations for the electron; in particular, the normalization condition will be of the same form (4.18) as before. Consequently, all characteristics of the positron will coincide with the ones of the electron; in particular, the positron spin will be $+\hbar/2$. However, inasmuch as the initial equations contained another sign in front of $A_\mu$, the charge and the direction of the magnetic moment will be opposite as compared with those of the electron.

Concluding the section let us discuss the behavior of the electron in external fields. At the end of Sec. 3 it was pointed out that in this situation $A_\mu$ is to be replaced by $A_\mu + A_\mu^{\text{ext}}$ in Eqs. (3.4)–(3.7) where $A_\mu^{\text{ext}}$ is the external four-potential. If the external field far exceeds the electronic self-field, one may put $A_\mu = v_\mu = 0$ in (3.4), which leads to the customary Dirac equation with one bispinor $\psi_1$. One may state that $\psi_2 = 0$ in this event in accord with Eq. (3.5) at $A_\mu = v_\mu = 0$. If there exists a physically reasonable (localized in space) solution for $\psi_1$, the solution can be regarded as a zeroth approximation and one may seek corrections to it in terms of expansions in powers of $\alpha$. So, for example, one may obtain the Lamb shift of levels in a hydrogen atom. If an external magnetic field is present in the equation for $\psi_1$, one can find the normal magnetic moment of the electron [2]. Upon taking the electronic self-field into consideration as a perturbation one is able to compute the anomalous part of the moment.

## 5. Physical meaning of fundamental constants

In all scientific papers and even in textbooks on QED one invariably puts $\hbar = c = 1$. This is rather surprising for the explicit writing of these constants, without essentially cluttering up formulae, may help in understanding the physical meaning of different terms. In all other branches of physics, when there are several dimensional constants, for the sake of convenience one introduces dimensionless quantities. We have done this in Eqs. (4.2) and (4.3). Somewhat unexpected is that the mass $m$ has disappeared off the equations as long as the fine-structure constant $\alpha = e^2/\hbar c$ that figures in them does not contain $m$. At the same time, one of the important procedures in QED is the renormalization of electron mass [2–3]. That is the reason why one retains the dimensional quantity $m$ upon setting $\hbar = c = 1$. A procedure cannot, however, disappear only because one has chosen other units. For this reason the procedure called the mass renormalization in QED has actually another meaning. In order to clarify the meaning we turn our attention to the fact that the mass does figure in (4.2)–(4.3) though implicitly, namely, in the adopted unit of length $\lambda\!\!\!^{-} = \hbar/mc$. Consequently, from the viewpoint of Eqs. (4.2)–(4.3) the renormalization of mass would correspond to a renormalization of unit of length. But the electron Compton wavelength $\lambda\!\!\!^{-}$ is in no way singled out as compared with other possible lengths and it would appear rather absurd to speak of its renormalization. In actual fact no renormalization of unit of length is involved but the essence of the procedure consists in representing spatial derivatives in terms of a series $\tilde{\nabla} = \tilde{\nabla}_0 + \tilde{\nabla}_1 + \tilde{\nabla}_2 + \ldots$. Such a representation of derivatives has long been known and was used in particular when solving the Boltzmann equation in the theory of gases. So, for example, in Ref. [12] one can meet an expansion of the type

$$\frac{\partial}{\partial t} = \frac{\partial_0}{\partial t} + \frac{\partial_1}{\partial t} + \frac{\partial_2}{\partial t} + \ldots .$$

Aside from the elucidation of the sense of the mass renormalization the foregoing leads to a more important conclusion. The fundamental meaning pertains not to the mass $m$ but to the length $\lambdabar = \hbar/mc$. Therefore the quantity $m$ in the Dirac equation must be replaced with $\hbar/\lambdabar c$. As a consequence, the correct form of Eq. (3.4), for instance, is

$$ic\hbar\gamma^{\mu}\frac{\partial\psi_1}{\partial x^{\mu}} - eA_{\mu}\gamma^{\mu}\psi_1 - \frac{\hbar c}{\lambdabar}\psi_1 - ev_{\mu}\gamma^{\mu}\psi_2 = 0\,. \tag{5.1}$$

One can imply in all other above equations that the letter $m$ denotes merely the combination $\hbar/\lambdabar c$ for brevity sake. One obtains the ordinary Dirac equation if one puts $v_{\mu} = 0$ in (5.1). The form of (5.1) is logical from the point of view of dimensions: seeing that the first term contains a derivative with respect to $x^{\mu}$, the third term should be supplemented with a factor having the dimension of $1/\lambdabar$.

We are now in position to elucidate the physical meaning of the constants $\lambdabar$, $c$, $e$ and $\hbar$ that enter into (5.1). First of all it should be emphasized that dimensional constants have no absolute meaning because their magnitude is determined only by comparing with a standard. We begin with $\lambdabar$. This constant is a standard of length for the Dirac equation. The standard is connected with the radius of the electron although it is not equal to the later according to Sec. 4. The next natural standard would be a standard of time $\tau_0$. However, as the second standard one utilizes the constant $c$ present in the Maxwell equations and equal to the speed of light in vacuum. If need be, one may find therefrom that $\tau_0 = \lambdabar/c = 1.3\cdot10^{-21}$ s. The next standard is a standard of charge $e$ equal to the charge of the electron. In Sec. 4 a hypothesis was advanced according to which the solution to equations describing the free electron exists only at the standard value of $\alpha$ ($\alpha \approx 1/137$). If the hypothesis proves to be correct, the fourth constant is not independent but is found from the relationship $\hbar = e^2/\alpha c$. It is precisely the value of $\hbar$ at which the electron exists. At the same time, the physical meaning of this constant consists in that it determines the magnitude of the electron spin.

It will be more logical to take $\lambdabar$, $c$ and $\hbar$ as the three basic constants. Then $\lambdabar$ will be a standard for translational motion whereas $\hbar$ for rotational one. The electron charge will be determined by the relationship $e^2 = \alpha c\hbar$. It is precisely the value of $e$ at which the electron exists. It is seen from the last formula that the electron exists only owing to the rotation ($\hbar \neq 0$).

It stems from Eq. (5.1) and the foregoing that the electron mass figures in no equations, the equations of motion inclusive, and thereby it is not a characteristic of the electron. We shall demonstrate nevertheless that the notion of the mass comes into being in the non-relativistic approximation, i.e., in the limit as $c \to \infty$. This limit should be taken with caution inasmuch as it is seen from the formula $\alpha = e^2/\hbar c$ that $\alpha \to 0$ when $c \to \infty$, that is to say, the electromagnetic interaction seemingly vanishes in this limit, which follows from (4.2) and (4.3). For this reason we introduce a parameter $m$ according to $m = \hbar/\lambdabar c$ and proceed to the limit $\lambdabar \to 0$ simultaneously with the limit $c \to \infty$ so that the parameter $m$ remain finite. However we have the limit $\tau_0 = \lambdabar/c \to 0$ concurrently with the limit $\lambdabar \to 0$. This means that we must pass to new units where the speed of light will be absent while the introduction of the parameter $m$ enables us to do this. So, for example, in the event of a hydrogen atom one utilizes atomic units [13]. Since $\alpha \to 0$, the action of the electromagnetic field of the electron upon the electron itself disappears. Consequently, we should put $A_\mu = v_\mu = 0$ in (5.1) in line with (4.2)–(4.3). Upon implying that the electron resides in an external field we write $A_\mu^{\text{ext}}$ instead of $A_\mu$ in (5.1) and replace the combination $\hbar/\lambdabar c$ in the third term by $m$. As a result we arrive at the usual Dirac equation for an electron in an external field. The subsequent passage to the limit $c \to \infty$ is carried out in a standard manner [2–3]. In a first approximation we obtain the Schrödinger equation from which it becomes clear by passing to classical mechanics that $m$ is the mass of the electron. It should be stressed that the above relationship $m = \hbar/\lambdabar c$ does not amount to asserting that the mass of the electron is equal to $\hbar/\lambdabar c$ seeing that the last combination was considered only when $c \to \infty$ and $\lambdabar \to 0$ whereas such a double limit is not unique. The limit will be unique if we require that the relation $\mathcal{E} = mc^2$ connecting the energy $\mathcal{E}$ and mass $m$ of a free particle at rest hold. In the case of the electron we have (4.27) for $\mathcal{E}$ from which

$$m = \frac{\hbar}{\lambdabar c}\left\{1 - \alpha c_0 + \alpha \int\limits_{(\infty)} \left[\frac{1}{8\pi}\left(\tilde{E}^2 + \tilde{H}^2\right) - \overline{\varphi}\left(\tilde{\psi}_1^* \tilde{\psi}_1 - \tilde{\psi}_2^* \tilde{\psi}_2\right)\right] d\tilde{V}\right\}. \tag{5.2}$$

Factually, this relationship slightly redefines $\lambdabar$ as long as the electron mass $m$ is known experimentally. As a consequence we see that the concept of the mass is a non-relativistic notion and one cannot speak of dependence of the mass upon its velocity because the rest mass and energy figure in $m = \mathcal{E}/c^2$. In this respect the author agrees with the assertions in Ref. [14].

The form of the Dirac equation as in (5.1) enables one to explain and calculate the mass spectrum of elementary particles which lacks reasonable explanation up till now. Equations (3.2), (3.4)–(3.9) (in which $m = \hbar/\lambdabar c$) can have not only the solution relevant to the electron (positron) but also solutions of higher energy $\mathcal{E} = \int T^{00} dV$ that must correspond to other leptons. Upon knowing the energy spectrum we shall obtain the mass spectrum by $m = \mathcal{E}/c^2$. It is quite possible that the solutions relevant to heavier leptons will not be strictly stationary but will prove to be metastable. When such is the case one will be able to calculate the lifetime as well.

The neutrinos may correspond to objects moving at the speed of light. Upon presuming that the motion occurs along the $z$-axis, in the simplest case the dependence of all functions upon coordinates and time is of the form $F = F(x, y, z - ct)$. Then $\partial F/\partial t = -c\partial F/\partial z$. In particular, Eqs. (4.2) and (4.3) acquire the form

$$-i\frac{\partial\tilde{\psi}_1}{\partial\tilde{z}} + i\boldsymbol{\alpha}\tilde{\nabla}\tilde{\psi}_1 + \alpha\tilde{\mathbf{A}}\boldsymbol{\alpha}\tilde{\psi}_1 - \alpha\tilde{\varphi}\tilde{\psi}_1 - \beta\tilde{\psi}_1 - \alpha\tilde{v}_0\tilde{\psi}_2 + \alpha\tilde{\mathbf{v}}\boldsymbol{\alpha}\tilde{\psi}_2 = 0\,, \tag{5.3}$$

$$-i\frac{\partial\tilde{\psi}_2}{\partial\tilde{z}} + i\boldsymbol{\alpha}\tilde{\nabla}\tilde{\psi}_2 + \alpha\tilde{\mathbf{A}}\boldsymbol{\alpha}\tilde{\psi}_2 - \alpha\tilde{\varphi}\tilde{\psi}_2 - \beta\tilde{\psi}_2 + \alpha\tilde{v}_0\tilde{\psi}_1 - \alpha\tilde{\mathbf{v}}\boldsymbol{\alpha}\tilde{\psi}_1 = 0\,. \tag{5.4}$$

Although there are electric and magnetic fields in the interior of the neutrino according to these equations, the neutrino as a whole must be neutral so that we should set $I = 0$ in (3.11). When moving the neutrino can evolve; in particular, one type of the neutrinos can periodically transform into another (the neutrino oscillations). To describe the neutrino evolution in its motion one should look for a solution to equations in the form $F = F(x, y, z - ct, t)$.

## 6. Concluding remarks

QED was constructed during two decades by the trial-and-error method. Its methods of calculations were frequently found empirically without leaning heavily upon mathematics although they finally gave extremely accurate predictions that were in excellent agreement with experiment. Even the basis of QED, the second quantization, was introduced by Dirac logically inconsistently. When considering assemblies of particles in his textbook [11] Dirac always takes as a ket for the assemblies the product of kets for each particle by itself $|a_1\rangle|b_2\rangle|c_3\rangle\ldots|g_n\rangle$ without mentioning anywhere that such a representation holds only for noninteracting particles. The fact that one is dealing with systems of noninteracting particles is clearly stated in §§ 64 and 65 of Ref. [13] devoted to the second quantization. Subsequently the apparatus of creation and annihilation operators strictly applicable only for noninteracting particles is tacitly carried over to systems of interacting particles. Moreover, the apparatus is extended to the coefficients $\hat{a}_s$ and $\hat{b}_s^+$ in expansions of the type (2.4) where one is dealing with only one particle (the electron). As a result it was empirically established that the second-quantization apparatus developed by Dirac without strict substantiation excellently works in QED. The essence of the method of the second quantization as a convenient mathematical tool in QED is discussed at the end of Sec. 2. It is worth remarking that the wave function of a system of $N$ interacting particles can always be strictly expanded in an infinite series in terms of products of one-particle functions (see, e.g., [15]). However, one can work with such an infinite series only approximately upon retaining a finite and small number of terms on condition that they describe the situation rather well. Such an approximate method called the second quantization as well can be used in a many-particle theory but it has nothing to do with the QED methods.

Analysis of QED equations enables one to establish that the electron must be described by two mutually connected bispinors rather than by one. In the paper two physical reasons are adduced which point to the necessity of this. First, as a result of the annihilation with a positron the electronic wave function disappears, which would be impossible with one bispinor. Secondly, in order to completely compensate the electric repulsive forces in the electron an amount of positive charge is required and therefore a second bispinor is needed to describe this. It should be remarked that, if a quartspinor is used instead of the two bispinors, the basic equations of the present paper can be recast in a form similar to the standard Dirac theory as shown in Appendix B.

Instead of $q$-numbers employed in QED which have no direct physical meaning, all functions that characterize the electron can be expressed in terms of $c$-numbers. It has been possible to reduce the starting equations whose number is equal to 16 to a system of 6 equations for 6 basic functions. Such a system can be solved only numerically, which is not a simple matter. A hypothesis was put forward according to which the system has a particle-like solution relevant to the electron only if the fine-structure constant $\alpha$ has the well-known experimental value ($\alpha \approx 1/137$). This would explain why $\alpha$ has just this value. The free electron is of a rather large size lying in the range between the electron Compton wavelength and the Bohr radius in a hydrogen atom, and resembles a cloud. It is discussed in Sec. 4 why not only does such a size of the electron not contradict experiment, but on the contrary it is supported by all experimental data concerning quantum mechanics.

Analysis of the physical meaning of the fundamental constants that figure in the Dirac equation leads to the conclusion that the equation must contain the electron Compton wavelength $\lambda\!\!\!^{-}$ rather than the electron mass $m$. As a consequence, the electron mass will not figure in any equations and thereby it is not a characteristic of the electron. The notion of the mass makes its appearance in the non-relativistic limit, which amounts to saying that the mass is a non-relativistic concept.

The form of the Dirac equation without $m$ as in (5.1) provides a possibility of considering not only the electron but other leptons as well. Consequently, one can calculate the mass spectrum of the leptons which lacks reasonable explanation up till now. Each lepton has its own energy $\mathcal{E}$ that can be found upon solving the equations obtained and thereupon the rest mass $m$ can be computed by $m = \mathcal{E}/c^2$. As to the neutrinos they may correspond to objects moving at the speed of light.

## Appendix A

Besides (4.52) the following equations can be obtained from (4.6)–(4.9):

$$\nabla^2 f_2^{(1)} - \frac{f_2^{(1)}}{\rho^2} - \nu^2 f_2^{(1)} + \rho z Q_2 = 0, \quad \nabla^2 f_3^{(1)} - \nu^2 f_3^{(1)} + z Q_3 = 0\,,$$

$$\nabla^2 f_4^{(1)} - \frac{f_4^{(1)}}{\rho^2} - \nu^2 f_4^{(1)} + \rho Q_4 = 0\,, \tag{A.1}$$

wherein

$$Q_2 = -\frac{\partial G_3^{(1)}}{\rho \partial \rho} + \frac{\partial G_4^{(1)}}{z \partial z} - (2 - \alpha c_0) G_2^{(1)}\,, \tag{A.2}$$

$$Q_3 = \rho \frac{\partial G_2^{(1)}}{\partial \rho} + \frac{\partial G_1^{(1)}}{z \partial z} + 2 G_2^{(1)} + \alpha c_0 G_3^{(1)}\,, \tag{A.3}$$

$$Q_4 = \frac{\partial G_1^{(1)}}{\rho \partial \rho} - z \frac{\partial G_2^{(1)}}{\partial z} - G_2^{(1)} + \alpha c_0 G_4^{(1)}\,. \tag{A.4}$$

It is worthy of remark that the combination of the type $\nabla^2 F - F/\rho^2$ that figures in (A.1) is the $\dot{\varphi}$-component of the Laplacian of a vector function $\mathbf{F}$ provided it does not depend on $\dot{\varphi}$. The same combination is present in (4.15).

By analogy with Eqs. (4.51) and (4.59) the equations of (A.1) may be rewritten in the integral form:

$$f_2^{(1)} = \frac{1}{4\pi} \int\limits_{(\infty)} \frac{e^{-\nu|\mathbf{r}-\mathbf{r}'|}}{|\mathbf{r}-\mathbf{r}'|} \cos(\dot{\varphi} - \dot{\varphi}') \rho' z' Q_2(\mathbf{r}') d\mathbf{r}'\,, \qquad f_3^{(1)} = \frac{1}{4\pi} \int\limits_{(\infty)} \frac{e^{-\nu|\mathbf{r}-\mathbf{r}'|}}{|\mathbf{r}-\mathbf{r}'|} z' Q_3(\mathbf{r}') d\mathbf{r}'\,,$$

$$f_4^{(1)} = \frac{1}{4\pi} \int\limits_{(\infty)} \frac{e^{-\nu|\mathbf{r}-\mathbf{r}'|}}{|\mathbf{r}-\mathbf{r}'|} \cos(\dot{\varphi} - \dot{\varphi}') \rho' Q_4(\mathbf{r}') d\mathbf{r}'\,. \tag{A.5}$$

It may be demonstrated that the functions $f_i^{(1)}$ as given by Eqs. (4.59) and (A.5) conform with their structure defined in (4.24).

When working with such integrals it is convenient to use the spherical coordinates $r'$, $\theta'$, $\dot{\varphi}'$ and Gegenbauer's addition theorem that enables one to express $\exp(-\nu|\mathbf{r} - \mathbf{r}'|) / |\mathbf{r} - \mathbf{r}'|$ in terms of a series that contains modified Bessel functions and Legendre polynomials [16] [if $\nu = 0$, this

gives 1/|**r** – **r**′| of (4.50) and (4.51)]. In this case the angular dependence of all functions in (4.59) and (A.5) can be represented in terms of expansions in the Legendre polynomials $P_k(\cos\theta)$ and in their derivatives.

## Appendix B

We rewrite Eq. (5.1) and Eq. (3.5) in the new form:

$$ic\hbar\gamma^{\mu}\frac{\partial\psi_1}{\partial x^{\mu}} - eA_{\mu}\gamma^{\mu}\psi_1 - \frac{\hbar c}{\lambda}\psi_1 - ev_{\mu}\gamma^{\mu}\psi_2 = 0\,, \tag{B.1}$$

$$ic\hbar\gamma^{\mu}\frac{\partial\psi_2}{\partial x^{\mu}} - eA_{\mu}\gamma^{\mu}\psi_2 - \frac{\hbar c}{\lambda}mc^2\psi_2 + ev_{\mu}\gamma^{\mu}\psi_1 = 0\,. \tag{B.2}$$

Instead of the two bispinors $\psi_1$ and $\psi_2$ we introduce a quartspinor $\Psi$ with eight components according to

$$\Psi = \begin{pmatrix} \psi_1 \\ \psi_2 \end{pmatrix}. \tag{B.3}$$

We introduce also the following 8×8 matrices

$$\Gamma^{\mu} = \begin{pmatrix} \gamma^{\mu} & 0 \\ 0 & \gamma^{\mu} \end{pmatrix}, \qquad \Delta^{\mu} = \begin{pmatrix} 0 & \gamma^{\mu} \\ -\gamma^{\mu} & 0 \end{pmatrix}. \tag{B.4}$$

If two matrices are similarly partitioned into smaller submatrices, one can multiply them by entering the submatrices as elements in the ordinary matrix-product formula. For this reason two Eqs. (B.1) and (B.2) can be recast as one equation

$$ic\hbar\Gamma^{\mu}\frac{\partial\Psi}{\partial x^{\mu}} - eA_{\mu}\Gamma^{\mu}\Psi - \frac{\hbar c}{\lambda}\Psi - ev_{\mu}\Delta^{\mu}\Psi = 0\,. \tag{B.5}$$

Properties of the matrices $\Gamma^{\mu}$ and $\Delta^{\mu}$ can be established with use made of the well-known properties of the matrices $\gamma^{\mu}$ [2]. In particular one has

$$\Gamma^{\mu}\Gamma^{\nu} + \Gamma^{\nu}\Gamma^{\mu} = 2g^{\mu\nu}, \qquad \Delta^{\mu}\Delta^{\nu} + \Delta^{\nu}\Delta^{\mu} = -2g^{\mu\nu}\,, \tag{B.6}$$

$$\Gamma^{\mu}\Delta^{\nu} + \Gamma^{\nu}\Delta^{\mu} = 2g^{\mu\nu}\mathrm{E}\,, \tag{B.7}$$

where the 8×8 matrix E is

$$\mathrm{E} = \begin{pmatrix} 0 & 1 \\ -1 & 0 \end{pmatrix}. \tag{B.8}$$

To rewrite the current density $j^{\mu}$ of (3.9) in terms of the quartspinor $\Psi$ we introduce a Dirac conjugate quartspinor $\overline{\Psi} = \Psi^{*}\Gamma^{0}$ and two 8×8 matrices

$$\mathrm{H}^{\mu} = \begin{pmatrix} \gamma^{\mu} & 0 \\ 0 & -\gamma^{\mu} \end{pmatrix}, \qquad \Theta = \begin{pmatrix} 1 & 0 \\ 0 & -1 \end{pmatrix} \tag{B.9}$$

with properties

$$\Gamma^{\mu}\Theta = \Theta\Gamma^{\mu} = \mathrm{H}^{\mu}, \qquad \Delta^{\mu}\Theta + \Theta\Delta^{\mu} = 0\,. \tag{B.10}$$

By the way, Eqs. (3.6) and (3.7) amount now to

$$ic\hbar\frac{\partial\overline{\Psi}}{\partial x^{\mu}}\Gamma^{\mu} + eA_{\mu}\overline{\Psi}\Gamma^{\mu} + \frac{\hbar c}{\lambda\!\!\!^{-}}\overline{\Psi} - ev_{\mu}\overline{\Psi}\Delta^{\mu} = 0\,. \tag{B.11}$$

The current density $j^{\mu}$ of (3.9) assumes now the form

$$j^{\mu} = \overline{\Psi}\mathrm{H}^{\mu}\Psi\,. \tag{B.12}$$

It follows from Eq. (3.8) that the quantity of (B.12) must satisfy the equation of continuity

$$\frac{\partial}{\partial x^{\mu}}\overline{\Psi}\mathrm{H}^{\mu}\Psi = 0\,. \tag{B.13}$$

Multiplying Eq. (B.5) on the left by $\overline{\Psi}\Theta$, and Eq. (B.11) on the right by $\Theta\Psi$ and adding the results with use made of (B.10), we ascertain that Eq. (B.13) holds.

The condition of (3.2) takes the form

$$\overline{\Psi}\Theta\Delta^{\mu}\Psi = 0\,. \tag{B.14}$$

In order to make sure that the condition does not contradict the equations of motion we write down extra relationships between the 8×8 matrices:

$$\Theta\mathrm{E} + \mathrm{E}\Theta = 0, \quad \Theta\mathrm{E}\Gamma^{\mu} = \Gamma^{\mu}\Theta\mathrm{E} = \Theta\Delta^{\mu}, \quad \Theta\mathrm{E}\Delta^{\mu} + \Delta^{\mu}\Theta\mathrm{E} = 0\,. \tag{B.15}$$

Now multiplying Eq. (B.5) on the left by $\overline{\Psi}\Theta\mathrm{E}$, and Eq. (B.11) on the right by $\Theta\mathrm{E}\Psi$ and adding the results with use made of (B.15), we obtain

$$\frac{\partial}{\partial x^{\mu}}\overline{\Psi}\Theta\Delta^{\mu}\Psi = 0\,. \tag{B.16}$$

We see that Eq. (B.14) is consistent with this last equation.

As a result, all basic equations of the present paper can be expressed in terms of the quartspinor $\Psi$.

We introduce also the following 8×8 matrices

$$\mathrm{A}^{k} = \Gamma^{0}\Gamma^{k} = \begin{pmatrix} \alpha^{k} & 0 \\ 0 & \alpha^{k} \end{pmatrix}, \qquad \mathrm{Z}^{k} = \Gamma^{0}\Delta^{k} = \begin{pmatrix} 0 & \alpha^{k} \\ -\alpha^{k} & 0 \end{pmatrix}, \tag{B.17}$$

where $k = 1, 2, 3$, and $\alpha^{k}$ are the components of the well-known matrix $\boldsymbol{\alpha}$ [2]. It is worth pointing out that

$$\mathrm{A}^{k}\mathrm{Z}^{l} + \mathrm{Z}^{l}\mathrm{A}^{k} = 2g^{kl}\mathrm{E}\,. \tag{B.18}$$

If we multiply Eq. (B.5) by $\Gamma^0$ with account taken of the fact that $(\Gamma^0)^2 = 1$ by (B.6), we shall have

$$i\hbar\frac{\partial\Psi}{\partial t} + ic\hbar \mathrm{A}^k \frac{\partial\Psi}{\partial x^k} - e\varphi\Psi - eA_k \mathrm{A}^k\Psi - \frac{c\hbar}{\lambda\!\!\!^{-}}\mathrm{B}\Psi - ev_0\mathrm{E}\Psi - ev_k \mathrm{Z}^k\Psi = 0 \qquad \text{(B.19)}$$

with $\mathrm{B} = \Gamma^0$. This equation can be rewritten as

$$i\hbar\frac{\partial\Psi}{\partial t} = H\Psi\,. \qquad \text{(B.20)}$$

Therefore the Hamiltonian $H$ in the present case is

$$H = c\mathrm{A}^k(\hat{p}_k + \frac{e}{c}A_k) + \frac{c\hbar}{\lambda\!\!\!^{-}}\mathrm{B} + e\varphi + ev_0\mathrm{E} + ev_k\mathrm{Z}^k\,, \qquad \text{(B.21)}$$

where

$$\hat{p}_k = -i\hbar\frac{\partial}{\partial x^k}\,. \qquad \text{(B.22)}$$

To compare this $H$ with the ordinary Dirac Hamiltonian it will be helpful to remark that $A_\mu = (\varphi, -\mathbf{A})$ and $v_\mu = (v_0, -\mathbf{v})$.